\RequirePackage{fix-cm}
\documentclass[smallextended]{svjour3}       % onecolumn (second format)
\smartqed  % flush right qed marks, e.g. at end of proof
\usepackage{graphicx}
\usepackage{amsfonts}
\usepackage{amssymb}
\usepackage{amsmath}
\newtheorem{Theorem}{Theorem}
\newtheorem{Proposition}{Proposition}

\begin{document}

\title{Dealing with the Inventory Risk\thanks{This research has been conducted within the Research Initiative ``Microstructure des March\'es Financiers'' under the aegis of the Europlace Institute of Finance.}}

\subtitle{A solution to the market making problem}

\author{Olivier Gu\'eant         \and
        Charles-Albert Lehalle \and
        Joaquin Fernandez-Tapia
}

\institute{Olivier Gu\'eant \at
              Universit\'e Paris-Diderot, UFR de Math\'ematiques, Laboratoire Jacques-Louis Lions. 175, rue du Chevaleret, 75013 Paris, France.\\
              \email{olivier.gueant@ann.jussieu.fr}\\
           \and
           Charles-Albert Lehalle\at
           Head of Quantitative Research, Cr\'edit Agricole Cheuvreux. 9, Quai du Pr\'esident Paul Doumer, 92400 Courbevoie, France.\\
           \email{clehalle@cheuvreux.com}\\
           \and
           Joaquin Fernandez-Tapia \at
           Universit\'e Pierre et Marie Curie, LPMA.
           4 place Jussieu, 75005 Paris, France.\\
}

\date{This draft: July 2012}
% The correct dates will be entered by the editor

\maketitle

\begin{abstract}
Market makers continuously set bid and ask quotes for the stocks they have under consideration. Hence they face a
complex optimization problem in which their return, based on the bid-ask spread they quote and the frequency at which they indeed provide liquidity, is challenged by the price risk they bear due to their inventory. In this paper, we consider a stochastic control problem similar to the one introduced by Ho and Stoll \cite{ho1981optimal} and formalized mathematically by Avellaneda and Stoikov \cite{avellaneda2008high}. The market is modeled using a reference price $S_t$ following a Brownian motion with standard deviation $\sigma$, arrival rates of buy or sell liquidity-consuming orders depend on the distance to the reference price $S_t$ and a market maker maximizes the expected utility of its P\&L over a finite time horizon. We show that the Hamilton-Jacobi-Bellman equations associated to the stochastic optimal control problem can be transformed into a system of linear ordinary differential equations and we solve the market making problem under inventory constraints. We also shed light on the asymptotic behavior of the optimal quotes and propose closed-form approximations based on a spectral characterization of the optimal quotes.\\

\end{abstract}

\section{Introduction}

From a quantitative viewpoint, market microstructure is a sequence of auction games between market participants. It implements the balance between
supply and demand, forming an equilibrium traded price to be used as reference for valuation. The rule of each auction game
(fixing auction, continuous auction, ...) is fixed by the firm operating each trading venue. Nevertheless, most of all trading mechanisms on
electronic markets rely on market participants sending orders to a ``queuing system'' where their open interests are consolidated as ``liquidity
provision'' or form transactions \cite{amihud1980deallership}. The efficiency of such a process relies on an adequate timing between buyers and
sellers, to avoid too many non-informative oscillations of the transaction price (for more details and modeling, see for example
\cite{citeulike:7621540}).\\

In practice, it is possible to provide liquidity to an impatient buyer (respectively seller) and maintain an inventory
until the arrival of the next impatient seller (respectively buyer). Market participants focused on this kind of liquidity-providing activity are
called ``market makers''. On one hand they are buying at the bid price and selling at the ask price they choose, making money out of this ``bid-ask
spread''. On the other hand, their inventory is exposed to price fluctuations mainly driven by the volatility of the market (see
\cite{amihud1986asset,benston1974determinants,cohen1981transaction,garman1976market,ho1984dealer,o1986microeconomics}).\\

The recent evolution of both technology and market regulation reshaped the nature of the interactions between market participants during continuous electronic auctions, one consequence being the emergence of ``high-frequency market makers'' who are said to be part of 70\% of the electronic trades in the US (40\% in the EU and 35\% in Japan) and have a massively passive (\emph{i.e.} liquidity-providing) behavior -- a typical balance between passive and aggressive orders for such market participants being around 80\% of passive interactions (see \cite{citeulike:8423311}).\\

From a mathematical modeling point of view, the market making problem corresponds to the choice of optimal quotes (\emph{i.e.} the bid and ask prices) that such agents provide to other market participants, taking into account their inventory limits and their risk constraints often represented by a utility function (see \cite{cohen1979market,hendershott2009price,ho1983dynamics,madhavan1993analysis,mildenstein1983optimal,roll1984simple}).\\

Avellaneda and Stoikov proposed, in a widely cited paper \cite{avellaneda2008high}, an innovative framework for ``market making in an order book''. In their approach, rooted to an old paper by Ho and Stoll \cite{ho1981optimal}, the market is modeled using a reference price or \emph{fair price} $S_t$ following a Brownian motion with standard deviation $\sigma$, and the arrival of a buy or sell liquidity-consuming order at a distance $\delta$ from the reference price $S_t$ is described by a point process with intensity $A\exp (-k \delta)$, $A$ and $k$ being two positive constants which characterize statistically the liquidity of the stock.\\

We consider the same model as in \cite{avellaneda2008high} -- adding inventory limits -- and we show, using a new change of variables, that the Hamilton-Jacobi-Bellman equations associated to the problem boil down to a system of linear ordinary differential equations. This new change of variables (i) simplifies the computation of a solution since numerical approximation of partial differential equations is now unnecessary, and (ii) allows to study the asymptotic behavior of the optimal quotes. In addition to these two contributions, we use results from spectral analysis to provide an approximation of the optimal quotes in closed-form. Finally, we provide in the case of our model with inventory limits a verification theorem that was absent from the original paper (the admissibility of the quotes obtained in the original Avellaneda-Stoikov model appears in fact to be an open problem!).\\

Since Avellaneda and Stoikov seminal paper, other authors have considered related market making models. Cartea, Jaimungal and Ricci \cite{cartea2011buy} consider a more sophisticated model inspired from the Avellaneda-Stoikov one\footnote{The objective function is different.}, including richer dynamics of market orders, impact on the limit order book, adverse selection effects and predictable $\alpha$. They obtained closed-form approximations of the optimal quotes using a first-order Taylor expansion. Cartea and Jaimungal \cite{cartea2012risk} recently used a similar model to introduce risk measures for high-frequency trading. Earlier, they used a model inspired from Avellaneda-Stoikov \cite{cartea2010modeling} in which the mid-price is modeled by a Hidden Markov Model. Guilbaud and Pham \cite{guilbaud2011optimal} also used a model inspired from the Avellaneda-Stoikov framework but including market orders and limit orders at best (and next to best) bid and ask together with stochastic spreads. Very recently, Guilbaud and Pham \cite{guilbaud2012optimal} used another model inspired from the Avellanada-Stoikov one in a pro-rata microstructure.\\

It is also noteworthy that the model we use to find the orders a market maker should optimally send to the market has been used in a totally different domain of algorithmic trading: optimal execution. Bayraktar and Ludkovski \cite{bayraktar2011liquidation} and Gu\'eant, Lehalle, Fernandez-Tapia \cite{GLFT} used indeed a similar model to optimally liquidate a portfolio.\\

This paper starts in section 2 with the description of the model. Section 3 is dedicated to the introduction of our change of variables and solves the control problem in the presence of inventory limits. Section 4 focuses on the asymptotic behavior of the optimal quotes and characterizes the asymptotic value using an eigenvalue problem that allows to propose a rather good approximation in closed-form. Section 5 generalizes the model in two different directions: (i) the introduction of a drift in the price dynamics and (ii) the introduction of market impact that may also be regarded as adverse selection. Section 6 carries out the comparative statics. Section 7 provides backtests of the model. Adaptations of our results are in use at Cheuvreux.

\section{Setup of the model}

Let us fix a probability space $(\Omega, \mathcal{F}, \mathbb{P})$ equipped with a filtration $(\mathcal{F}_t)_{t\geq 0}$ satisfying the usual
conditions. We assume that all random variables and stochastic processes are defined on $(\Omega, \mathcal{F},(\mathcal{F}_t)_{t\geq 0},
\mathbb{P})$.\\

We consider a high-frequency market maker operating on a single stock\footnote{We suppose that this high-frequency market maker does not
``make'' the price in the sense that he has no market power. In other words, we assume that the size of his orders is small enough to consider price dynamics
exogenous. Market impact will be introduced in section 5.}. We suppose that the mid-price of this stock or more generally a reference price\footnote{This reference price for the stock can be thought of as a
smoothed mid-price for instance.} of the stock moves as an arithmetic Brownian motion\footnote{Since we will only consider short horizon problems, this assumption is almost equivalent to the usual Black-Scholes one.}: $$dS_t = \sigma dW_t$$

The market maker under consideration will continuously propose bid and ask prices denoted respectively $S^b_t$ and $S^a_t$ and will hence buy and sell shares according to the rate of arrival of market orders at the quoted prices. His inventory $q$, that is the (signed) quantity of shares he holds, is given by $$q_t = N^b_t - N^a_t$$ where $N^b$ and $N^a$ are the point processes (independent of $(W_t)_t$) giving the number of shares the market maker respectively bought and sold (we assume that transactions are of constant size, scaled\footnote{The only important hypothesis is the constant size of orders since we can easily replace $1$ by any positive size $\Delta$.} to $1$). Arrival rates obviously depend on the prices $S^b_t$ and $S^a_t$ quoted by the market maker and we assume, in accordance with the model proposed by Avellaneda and Stoikov \cite{avellaneda2008high}, that intensities $\lambda^b$ and $\lambda^a$ associated respectively to $N^b$ and $N^a$ depend on the difference between the quoted prices and the reference price (\emph{i.e.} $\delta_t^b = S_t - S^b_t$ and $\delta_t^a = S^a_t - S_t$) and are of the following form\footnote{Some authors also used a linear form for the intensity functions -- see \cite{ho1981optimal} for instance.}: $$\lambda^b(\delta^b) = A e^{-k \delta^b} = A \exp(-k(s-s^b)) \qquad \lambda^a(\delta^a) = A e^{-k \delta^a} = A \exp(-k(s^a-s))$$
where $A$ and $k$ are positive constants that characterize the liquidity of the stock. In particular, this specification means -- for positive $\delta^b$ and $\delta^a$ -- that the closer to the reference price an order is posted, the faster it will be executed.\\

As a consequence of his trades, the market maker has an amount of cash evolving according to the following dynamics: $$dX_t = (S_t + \delta^a_t) dN^a_t - (S_t -
\delta^b_t) dN^b_t$$

To this original setting introduced by Avellaneda and Stoikov (itself following partially Ho and Stoll \cite{ho1981optimal}), we add a bound $Q$ to the
inventory that a market maker is authorized to have. In other words, we assume that a market maker with inventory $Q$ ($Q>0$ depending in practice on
risk limits) will never set a bid quote and symmetrically that a market maker with inventory $-Q$, that is a short position of $Q$ shares in the
stock under consideration, will never set an ask quote. This realistic restriction may be read as a risk limit and allows to solve rigorously the problem.\\

Now, coming to the objective function, the market maker has a time horizon $T$ and his goal is to optimize the expected utility
of his P\&L at time $T$. In line with \cite{avellaneda2008high}, we will focus on CARA utility functions and we suppose that the market maker
optimizes:

$$\sup_{(\delta_t^a)_t,(\delta_t^b)_t \in \mathcal{A}} \mathbb{E}\left[-\exp\left(-\gamma (X_T + q_T S_T) \right)\right]$$ where $\mathcal{A}$ is
the set of predictable processes bounded from below, $\gamma$ is the absolute risk aversion coefficient characterizing the market maker, $X_T$ is the amount of
cash at time $T$ and $q_T S_T$ is the evaluation of the (signed) remaining quantity of shares in the inventory at time $T$ (liquidation at the reference
price $S_T$\footnote{Our results would be \emph{mutatis mutandis} the same if we added a penalization term $- b(|q_T|)$ for the shares remaining at
time $T$. The rationale underlying this point is that price risk prevents the trader from having important exposure to the stock. Hence, $q_t$ should naturally mean-revert around $0$.}).

\vspace{0.5cm}
\section{Characterization of the optimal quotes}

The optimization problem set up in the preceding section can be solved using the classical tools of stochastic optimal control. The first step of our reasoning is therefore to introduce the Hamilton-Jacobi-Bellman (HJB) equation associated to the problem. More exactly, we introduce a system of Hamilton-Jacobi-Bellman partial differential equations which consists of the following equations indexed by $q \in \lbrace -Q, \ldots, Q \rbrace$ for $(t,s,x) \in [0,T]\times \mathbb{R}^2$:\\

\noindent For $|q| < Q$:
$$\partial_t u(t,x,q,s) + \frac 12 \sigma^2 \partial^2_{ss} u(t,x,q,s)$$$$ + \sup_{\delta^b}
\lambda^b(\delta^b) \left[u(t,x-s+\delta^b,q+1,s) - u(t,x,q,s) \right]$$$$ + \sup_{\delta^a} \lambda^a(\delta^a) \left[u(t,x+s+\delta^a,q-1,s) -
u(t,x,q,s) \right] = 0$$
For $q=Q$:
$$\partial_t u(t,x,Q,s) + \frac 12 \sigma^2 \partial^2_{ss} u(t,x,Q,s)$$$$ + \sup_{\delta^a} \lambda^a(\delta^a) \left[u(t,x+s+\delta^a,Q-1,s) -
u(t,x,Q,s) \right] = 0$$
For $q=-Q$:
$$\partial_t u(t,x,-Q,s) + \frac 12 \sigma^2 \partial^2_{ss} u(t,x,-Q,s)$$$$ + \sup_{\delta^b} \lambda^b(\delta^b)
\left[u(t,x-s+\delta^b,-Q+1,s) - u(t,x,-Q,s) \right] = 0$$ with the final condition: $$\forall q \in \lbrace -Q, \ldots, Q \rbrace, \qquad
u(T,x,q,s) = -\exp\left(-\gamma(x+qs)\right)$$

To solve these equations we will use a change of variables based on two different ideas. First, the choice of a CARA utility function allows to
factor out the Mark-to-Market value of the portfolio ($x+qs$). Then, the exponential decay for the intensity functions $\lambda^b$ and $\lambda^a$
allows to reduce the Hamilton-Jacobi-Bellman (HJB) equations associated to our control problem to a linear system of ordinary differential equations:\\

\begin{Proposition}[Change of variables for (HJB)] Let us consider a family $(v_q)_{|q|\le Q}$ of positive functions solution of: $$\forall
q \in \lbrace -Q+1, \ldots, Q-1 \rbrace, \quad \dot{v}_q(t) = \alpha q^2 v_q(t) - \eta \left( v_{q-1}(t) + v_{q+1}(t) \right)$$ $$ \dot{v}_Q(t) =
\alpha Q^2 v_Q(t) - \eta v_{Q-1}(t)$$ $$\dot{v}_{-Q}(t) = \alpha Q^2 v_{-Q}(t) - \eta v_{-Q+1}(t)$$ with $\forall q \in \lbrace -Q, \ldots, Q
\rbrace,  v_q(T) = 1$, where $\alpha = \frac k2 \gamma \sigma^2$ and $\eta = A(1+\frac \gamma k)^{-(1+\frac k\gamma)}$.\\ Then, $u(t,x,q,s) =
-\exp(-\gamma(x+qs)){v_q(t)}^{-\frac \gamma k}$ is solution of (HJB).\\
\end{Proposition}

Then, the following proposition proves that there exists such a family of positive functions:

\begin{Proposition}[Solution of the ordinary differential equations] Let us introduce the matrix $M$ defined by:

\[M=
 \begin{pmatrix}
  \alpha Q^2 & -\eta & 0               & \cdots & \cdots            & \cdots           & 0 \\
  -\eta & \alpha (Q-1)^2 & -\eta & 0       & \ddots            & \ddots           & \vdots \\
  0              &    \ddots           & \ddots         & \ddots & \ddots           &   \ddots               & \vdots \\
    \vdots              &    \ddots           & \ddots         & \ddots & \ddots           &   \ddots               & \vdots \\
      \vdots              &    \ddots           & \ddots         & \ddots & \ddots           &   \ddots               & 0 \\
                    \vdots& \ddots         & \ddots          & 0       & -\eta & \alpha (Q-1)^2   & -\eta\\
 0                   & \cdots         & \cdots         & \cdots  & 0                & -\eta & \alpha Q^2
\end{pmatrix} \]

where $\alpha = \frac k2 \gamma \sigma^2$ and $\eta = A(1+\frac \gamma k)^{-(1+\frac k\gamma)}$.\\

Let us define $$v(t) = (v_{-Q}(t), v_{-Q+1}(t), \dots,  v_0(t), \dots, v_{Q-1}(t), v_{Q}(t))'$$$$ = \exp(-M(T-t)) \times (1, \dots, 1)'$$ Then,
$(v_q)_{|q|\le Q}$ is a family of positive functions solution of the equations of Proposition 1.\\

\end{Proposition}

Using the above change of variables and a verification approach, we are now able to solve the stochastic control problem, that is to find the value function of the problem and the optimal quotes:

\begin{Theorem}[Solution of the control problem] Let consider $(v_q)_{|q|\le Q}$ as in Proposition 2.\\

Then $u(t,x,q,s) = -\exp(-\gamma(x+qs)){v_q(t)}^{-\frac \gamma k}$ is the value function of the control problem.\\

Moreover, the optimal quotes are given by: $$ s - s^{b*}(t,q,s) =  \delta^{b*}(t,q) =
\frac{1}{k}\ln\left(\frac{v_{q}(t)}{v_{q+1}(t)}\right) + \frac 1\gamma \ln\left(1+\frac \gamma k\right), \quad  q\neq Q$$ $$ s^{a*}(t,q,s) - s = \delta^{a*}(t,q) =  \frac{1}{k}\ln\left(\frac{v_{q}(t)}{v_{q-1}(t)}\right) + \frac 1\gamma
\ln\left(1+\frac \gamma k\right), \quad q\neq -Q$$ and the resulting bid-ask spread quoted by the market maker is given by: $$\psi^*(t,q) = -\frac{1}{k}\ln\left(\frac{v_{q+1}(t)v_{q-1}(t)}{v_q(t)^2}\right) + \frac 2\gamma \ln\left(1+\frac \gamma
k\right), \quad |q| \neq Q$$

\end{Theorem}
\vspace{0.5cm}

\section{Asymptotic behavior and approximation of the optimal quotes}

\begin{figure}[!h] \center
  % Requires \usepackage{graphicx}
  \includegraphics[width=0.85\textwidth]{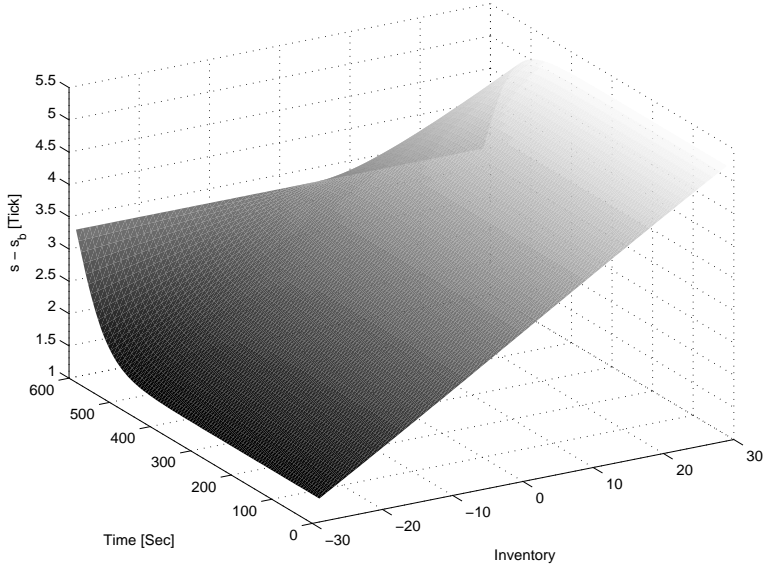}
 \caption{Behavior of the optimal bid quotes with time and inventory. $\sigma = 0.3\quad \mathrm{Tick}\cdot\mathrm{s}^{-1/2}$, $A = 0.9\quad\mathrm{s}^{-1}$, $k = 0.3\quad\mathrm{Tick}^{-1}$, $\gamma =
 0.01\quad\mathrm{Tick}^{-1}$, $T = 600\quad s$.}
 \label{ex1}
\end{figure}

\begin{figure}[!h] \center
  % Requires \usepackage{graphicx}
 \includegraphics[width=0.85\textwidth]{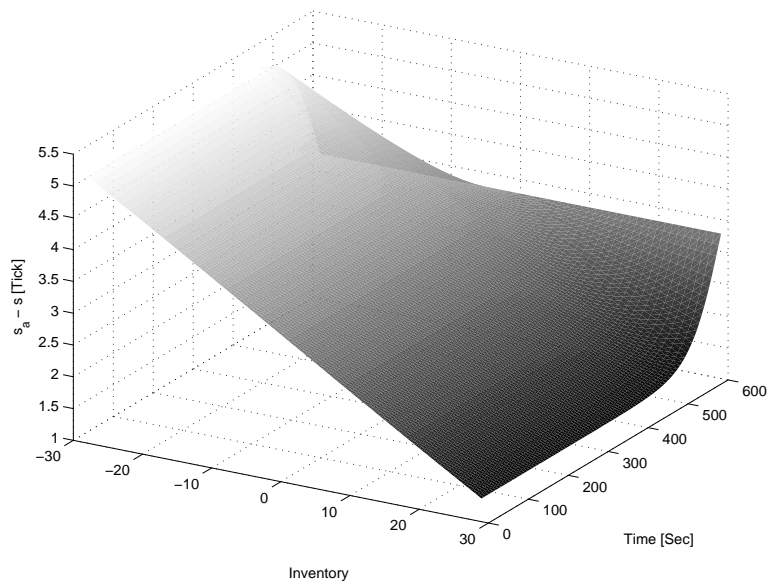}
 \caption{Behavior of the optimal ask quotes with time and inventory. $\sigma = 0.3\quad \mathrm{Tick}\cdot\mathrm{s}^{-1/2}$, $A = 0.9\quad\mathrm{s}^{-1}$, $k = 0.3\quad\mathrm{Tick}^{-1}$, $\gamma =
 0.01\quad\mathrm{Tick}^{-1}$, $T = 600\quad s$.}
 \label{ex1b}
\end{figure}

\begin{figure}[!h]
\center
\includegraphics[width=0.85\textwidth]{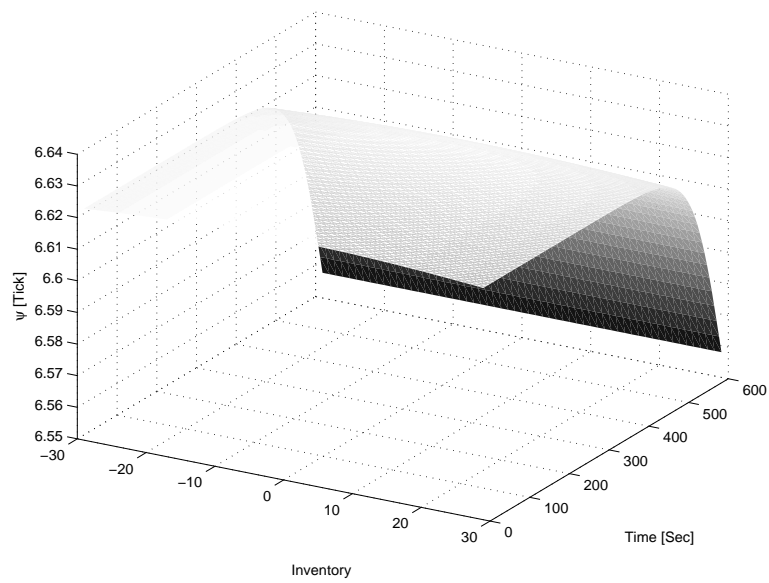}
\caption{Behavior of the resulting bid-ask spread with time and inventory. $\sigma = 0.3\quad \mathrm{Tick}\cdot\mathrm{s}^{-1/2}$, $A =
0.9\quad\mathrm{s}^{-1}$, $k = 0.3\quad\mathrm{Tick}^{-1}$, $\gamma = 0.01\quad\mathrm{Tick}^{-1}$, $T = 600\quad s$.}
\label{ex2}
\end{figure}

To exemplify our findings and in order to motivate the asymptotic approximations we shall provide, we plotted on Figure \ref{ex1} and Figure \ref{ex1b} the
behavior with time and inventory of the optimal quotes. The resulting bid-ask spread quoted by the market maker is plotted on Figure \ref{ex2}.\\

We clearly see that the optimal quotes are almost independent of $t$, as soon as $t$ is far from the terminal time $T$. This observation is at odds with the approximations proposed\footnote{In \cite{avellaneda2008high}, the approximations obtained by the authors using an expansion in $q$ leads to the following expressions:
$$\delta^{b*}_t \simeq \frac 1\gamma \ln\left(1+\frac \gamma k\right) + \frac{1+2q}{2}\gamma \sigma^2 (T-t)$$
$$\delta^{a*}_t \simeq \frac 1\gamma \ln\left(1+\frac \gamma k\right) + \frac{1-2q}{2}\gamma \sigma^2 (T-t)$$
One can easily show, using the results of Theorem 1, that these approximations are nothing but the Taylor expansions of the optimal quotes for $t$ close to $T$.} in Avellaneda and Stoikov \cite{avellaneda2008high} using an expansion in $q$. It motivates however the study of the asymptotic behavior of the quotes.\\

\vfill
\vspace{2cm}
\vfill

\begin{Theorem}[Asymptotics for the optimal quotes] The optimal quotes have asymptotic limits

$$\lim_{T\to+\infty} \delta^{b*}(0,q) = \delta^{b*}_\infty(q)$$ $$\lim_{T\to+\infty} \delta^{a*}(0,q) = \delta^{a*}_\infty(q)$$

that can be expressed as:

$$\delta^{b*}_\infty(q) =  \frac 1\gamma \ln\left(1+\frac \gamma k\right) + \frac 1k \ln\left(\frac{f^0_{q}}{f^0_{q+1}}\right) \quad
\delta^{a*}_\infty(q) =  \frac 1\gamma \ln\left(1+\frac \gamma k\right) + \frac 1k \ln\left(\frac{f^0_{q}}{f^0_{q-1}}\right)$$ where $f^0 \in
\mathbb{R}^{2Q+1}$ is an eigenvector corresponding to the smallest eigenvalue of the matrix $M$ introduced in Proposition 2 and characterized (up to a multiplicative constant) by:
$$f^0 \in \underset{f \in \mathbb{R}^{2Q+1}, \|f\|_2 = 1}{\mathrm{argmin}} \sum_{q=-Q}^Q \alpha q^2 {f_q}^2 + \eta \sum_{q=-Q}^{Q-1} (f_{q+1} - f_q)^2 + \eta {f_Q}^2 + \eta {f_{-Q}}^2$$
The resulting
bid-ask spread quoted by the market maker is asymptotically: $$\psi^*_\infty(q) =-\frac{1}{k}\ln\left(\frac{f^0_{q+1}f^0_{q-1}}{{f^0_q}^2}\right) +
\frac 2\gamma \ln\left(1+\frac \gamma k\right)$$ \end{Theorem}

The above result, along with the example of Figure \ref{ex1}, Figure \ref{ex1b} and Figure \ref{ex2}, encourages to approximate the optimal quotes and the resulting bid-ask spread by their asymptotic value. These asymptotic values depend on $f^0$ and we shall provide a closed-form approximation for $f^0$.\\

The above characterization of $f^0$ corresponds to an eigenvalue problem in $\mathbb{R}^{2Q+1}$ and we propose to replace it by a similar eigenvalue problem in $L^2(\mathbb{R})$ for which a closed-form solution can be computed. More precisely we replace the criterion

$$f^0 \in \underset{f \in \mathbb{R}^{2Q+1}, \|f\|_2 = 1}{\mathrm{argmin}} \sum_{q=-Q}^Q \alpha q^2 {f_q}^2 + \eta \sum_{q=-Q}^{Q-1} (f_{q+1} - f_q)^2 + \eta {f_Q}^2 + \eta {f_{-Q}}^2$$

by the following criterion for $\tilde{f}^0 \in L^2(\mathbb{R})$:

$$\tilde{f}^0 \in \underset{\|\tilde{f}\|_{L^2(\mathbb{R})}=1}{\mathrm{argmin}}  \int_{-\infty}^{+\infty} \left(\alpha x^2 \tilde{f}(x)^2 + \eta \tilde{f}'(x)^2\right) dx$$

The introduction of this new criterion is rooted to the following proposition that states (up to its sign) the expression for $\tilde{f}^0$ in closed form:

\begin{Proposition}
Let us consider $$\tilde{f}^0 \in \underset{\|\tilde{f}\|_{L^2(\mathbb{R})}=1}{\mathrm{argmin}} \int_{\mathbb{R}}{ \left(\alpha x^2 \tilde{f}(x)^2 + \eta \tilde{f}'(x)^2\right) dx}$$
Then: $$ \tilde{f}^0(x) = \pm \frac {1}{\pi^{\frac 14}}\left(\frac \alpha \eta\right)^{\frac 18}\exp\left(-\frac{1}{2}\sqrt{\frac{\alpha}{\eta}}x^2\right)$$

\end{Proposition}

From the above proposition, we expect $f_q^0$ to behave, up to a multiplicative constant, as  $\exp\left(-\frac{1}{2}\sqrt{\frac{\alpha}{\eta}}q^2\right)$. This heuristic viewpoint induces an approximation of the optimal quotes and the resulting optimal bid-ask-spread:

\begin{eqnarray*}
\delta^{b*}_\infty(q) &\simeq&  \frac 1\gamma \ln\left(1+\frac \gamma k\right) + \frac {1}{2k} \sqrt{\frac{\alpha}{\eta}} (2q+1)\\
 &\simeq& \frac 1\gamma \ln\left(1+\frac \gamma k\right) + \frac{2q+1}{2} \sqrt{\frac{\sigma^2\gamma}{2kA}\left(1+\frac{\gamma}{k}\right)^{1+\frac{k}{\gamma}}}
\\
\delta^{a*}_\infty(q) &\simeq&  \frac 1\gamma \ln\left(1+\frac \gamma k\right) - \frac {1}{2k} \sqrt{\frac{\alpha}{\eta}} (2q-1)\\
&\simeq& \frac 1\gamma \ln\left(1+\frac \gamma k\right) - \frac{2q-1}{2} \sqrt{\frac{\sigma^2\gamma}{2kA}\left(1+\frac{\gamma}{k}\right)^{1+\frac{k}{\gamma}}}
\end{eqnarray*}

$$\psi^*_\infty(q) \simeq \frac 2\gamma \ln\left(1+\frac \gamma k\right)+
\sqrt{\frac{\sigma^2\gamma}{2kA}\left(1+\frac{\gamma}{k}\right)^{1+\frac{k}{\gamma}}}$$

\begin{figure}[!h] \center
  % Requires \usepackage{graphicx}
  \includegraphics[width=0.45\textwidth]{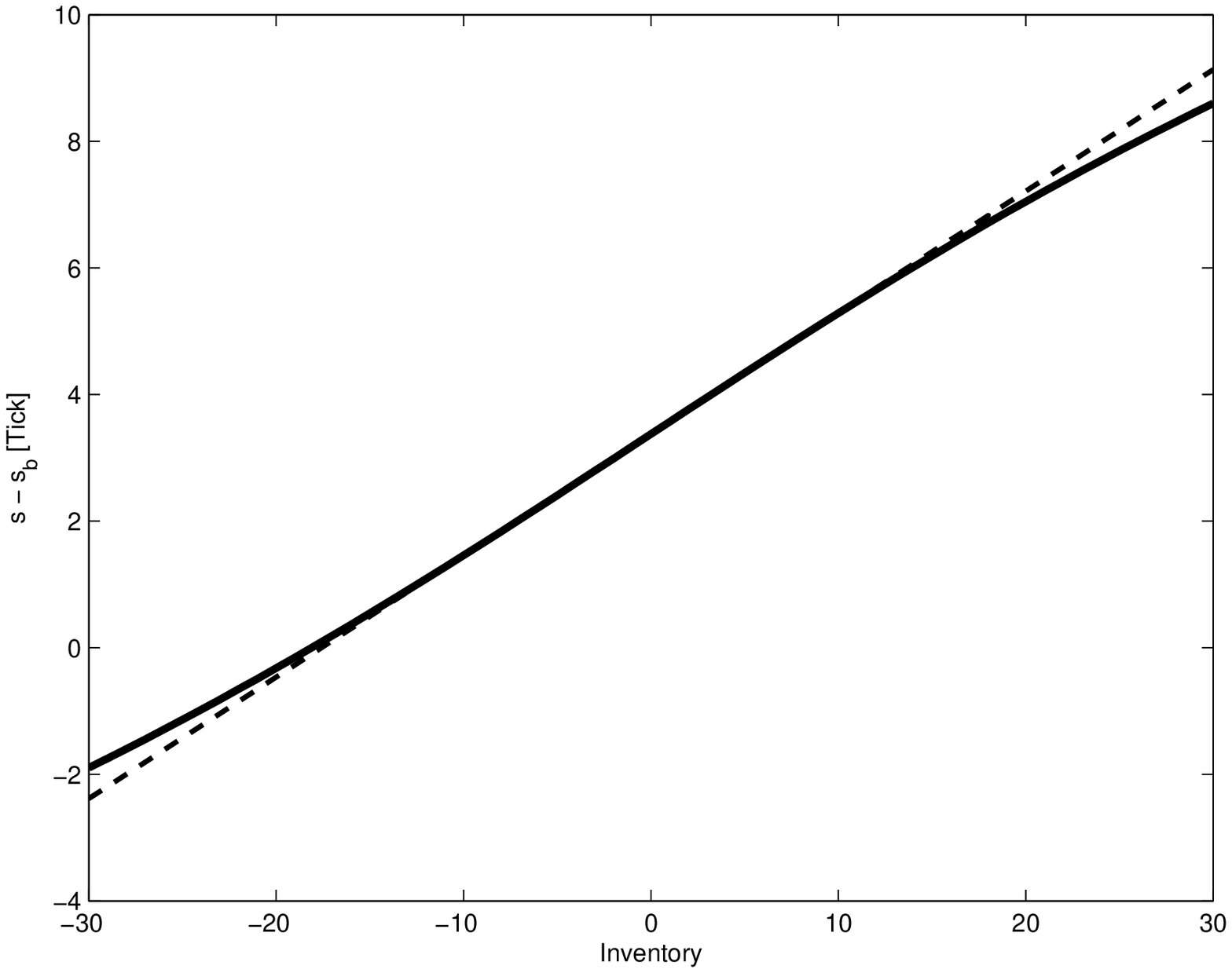}
  \hspace{0.1cm}
  \includegraphics[width=0.45\textwidth]{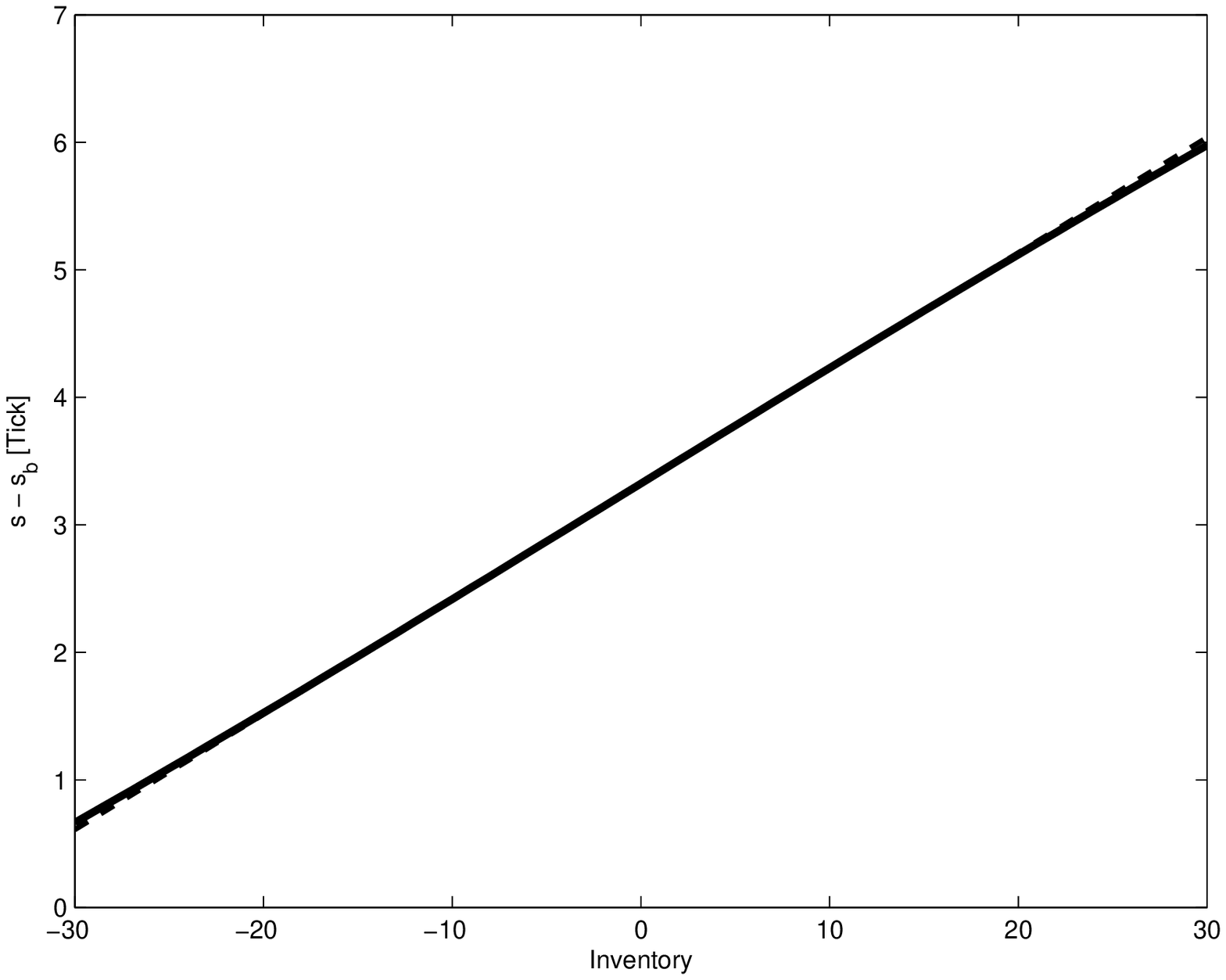}
  \label{ap}
    \caption{Asymptotic behavior of optimal bid quote (bold line). Approximation (dotted line). Left: $\sigma = 0.4\quad\mathrm{
  Tick}\cdot\mathrm{s}^{-1/2}$, $A = 0.9\quad\mathrm{s}^{-1}$, $k = 0.3\quad\mathrm{Tick}^{-1}$, $\gamma = 0.01\quad \mathrm{Tick}^{-1}$, $T =
  600\quad s$. Right: $\sigma = 1.0\quad\mathrm{ Tick}\cdot\mathrm{s}^{-1/2}$, $A = 0.2\quad\mathrm{s}^{-1}$, $k = 0.3\quad\mathrm{Tick}^{-1}$,
  $\gamma = 0.01\quad \mathrm{Tick}^{-1}$, $T = 600\quad s$.}
 \end{figure}

 \begin{figure}[!h]
\center
 \includegraphics[width=0.45\textwidth]{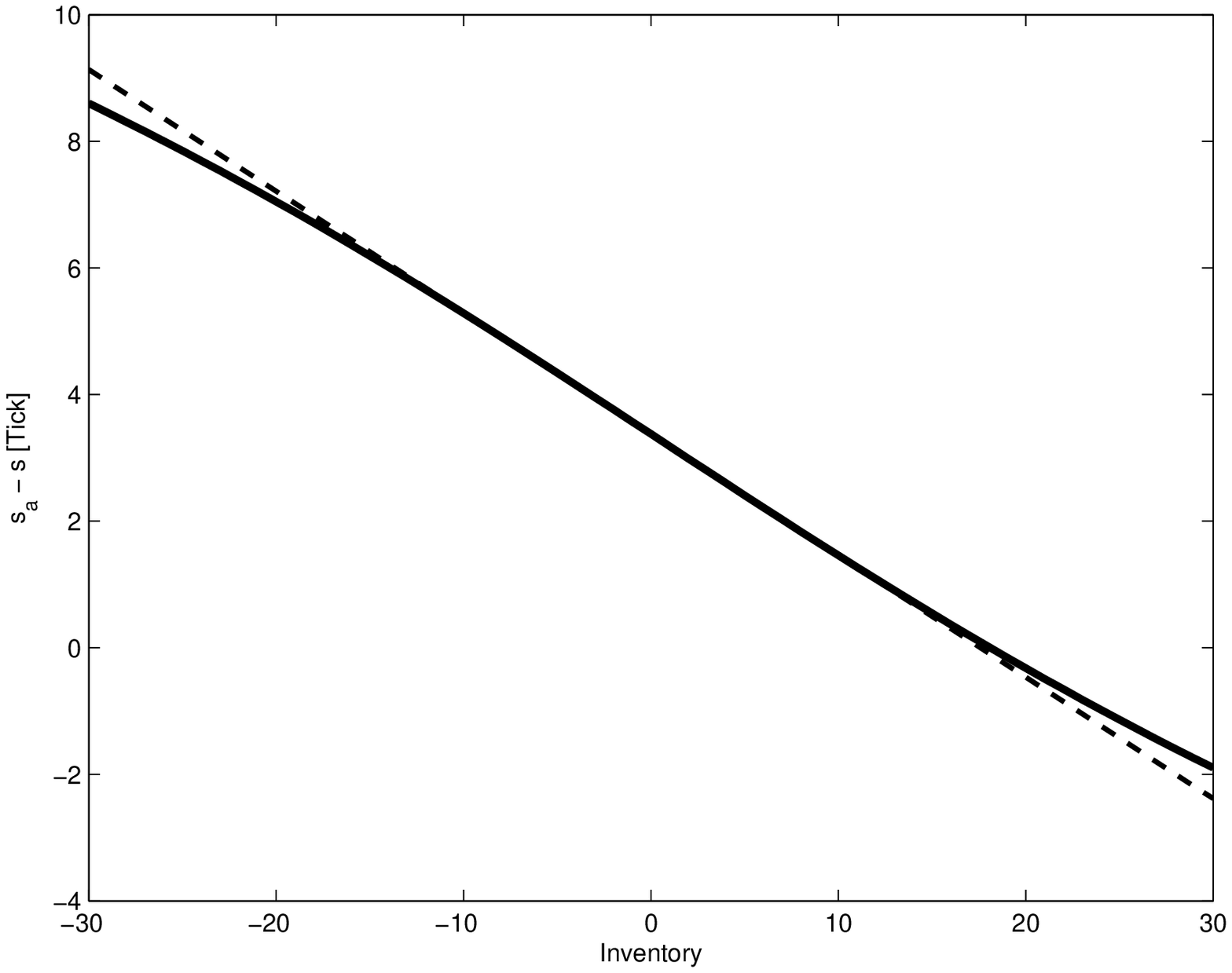}
 \hspace{0.1cm}
 \includegraphics[width=0.45\textwidth]{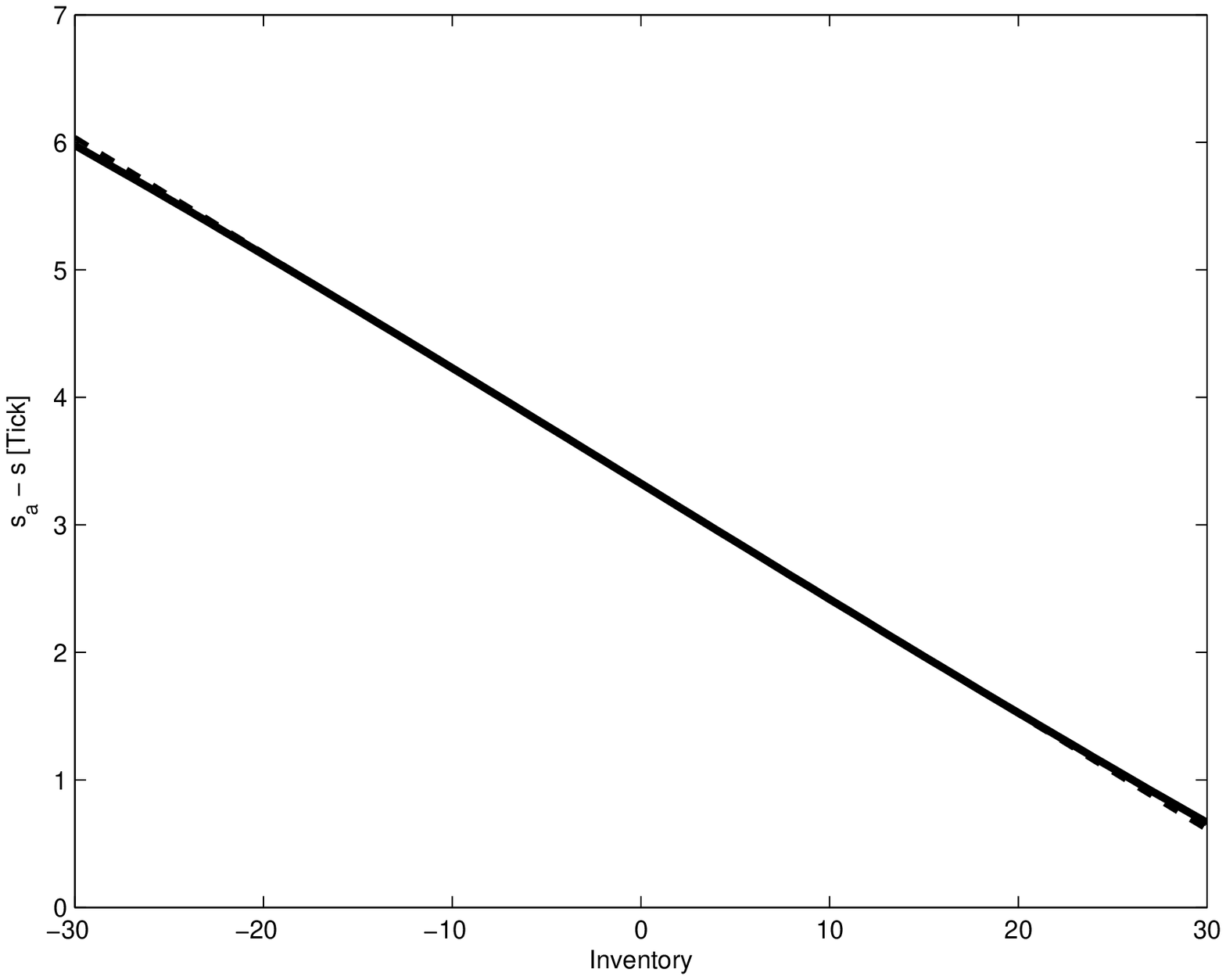}
  \label{ap2}
  \caption{Asymptotic behavior of optimal ask quote (bold line). Approximation (dotted line). Left: $\sigma = 0.4\quad\mathrm{
  Tick}\cdot\mathrm{s}^{-1/2}$, $A = 0.9\quad\mathrm{s}^{-1}$, $k = 0.3\quad\mathrm{Tick}^{-1}$, $\gamma = 0.01\quad \mathrm{Tick}^{-1}$, $T =
  600\quad s$. Right: $\sigma = 1.0\quad\mathrm{ Tick}\cdot\mathrm{s}^{-1/2}$, $A = 0.2\quad\mathrm{s}^{-1}$, $k = 0.3\quad\mathrm{Tick}^{-1}$,
  $\gamma = 0.01\quad \mathrm{Tick}^{-1}$, $T = 600\quad s$.}
\end{figure}

We exhibit on Figure \ref{ap} and Figure 5 the values of the optimal quotes, along with their associated approximations. Empirically, these approximations for the quotes are satisfactory in most cases and are always very good for small values of the inventory $q$. In fact, even though $f^0$ appears to be well approximated by the gaussian approximation, we cannot expect a very good fit for the quotes when $q$ is large because we are approximating expressions that depend on ratios of the form $\frac{f^0_q}{f^0_{q+1}}$ or $\frac{f^0_q}{f^0_{q-1}}$.\\

\section{Extensions of the model}

\subsection{The case of a trend in the price dynamics}

So far, the reference price was supposed to be a Brownian motion. In what follows we extend the model to the case of a trend in the price dynamics:
 $$dS_t = \mu dt + \sigma dW_t$$
In that case we have the following proposition (the proof is not repeated):

\begin{Proposition}[Solution with a drift]
Let us consider a family of functions $(v_q)_{|q|\le Q}$ solution of the linear system of ODEs that follows:
$$\forall q\in \lbrace -Q+1, \ldots, Q-1 \rbrace, \dot{v}_q(t) = (\alpha q^2 - \beta q) v_q(t) - \eta  \left( v_{q-1}(t) + v_{q+1}(t) \right)$$
$$ \dot{v}_Q(t) = (\alpha Q^2 - \beta Q) v_Q(t) - \eta v_{Q-1}(t)$$
$$\dot{v}_{-Q}(t) = (\alpha Q^2 + \beta Q) v_{-Q}(t) - \eta v_{-Q+1}(t)$$
with $\forall q \in \lbrace -Q, \ldots, Q \rbrace,  v_q(T) = 1$, where $\alpha = \frac k2 \gamma \sigma^2$, $\beta = k\mu$ and $\eta = A(1+\frac \gamma k)^{-(1+\frac k\gamma)}$.\\

Then, $u(t,x,q,s) = -\exp(-\gamma(x+qs)){v_q(t)}^{-\frac \gamma k}$ is the value function of the control problem.\\

The optimal quotes are given by:
$$s - s^{b*}(t,q,s) = \delta^{b*}(t,q) = \frac{1}{k}\ln\left(\frac{v_{q}(t)}{v_{q+1}(t)}\right) + \frac 1\gamma \ln\left(1+\frac \gamma k\right)$$
$$s^{a*}(t,q,s) - s = \delta^{a*}(t,q) = \frac{1}{k}\ln\left(\frac{v_{q}(t)}{v_{q-1}(t)}\right) + \frac 1\gamma \ln\left(1+\frac \gamma k\right)$$
and the resulting bid-ask spread of the market maker is :
$$\psi^*(t,q) = -\frac{1}{k}\ln\left(\frac{v_{q+1}(t)v_{q-1}(t)}{v_q(t)^2}\right) + \frac 2\gamma \ln\left(1+\frac \gamma k\right)$$
Moreover,

$$\lim_{T\to+\infty} \delta^{b*}(0,q) = \frac 1\gamma \ln\left(1+\frac \gamma k\right) + \frac 1k \ln\left(\frac{f^0_{q}}{f^0_{q+1}}\right)$$
$$\lim_{T\to+\infty} \delta^{a*}(0,q) = \frac 1\gamma \ln\left(1+\frac \gamma k\right) + \frac 1k \ln\left(\frac{f^0_{q}}{f^0_{q-1}}\right)$$
$$\lim_{T\to+\infty} \psi^*(0,q) = -\frac{1}{k}\ln\left(\frac{f^0_{q+1}f^0_{q-1}}{{f^0_q}^2}\right) + \frac 2\gamma \ln\left(1+\frac \gamma k\right)$$
where $f^0$ is an eigenvector corresponding to the smallest eigenvalue of:
\[
 \begin{pmatrix}
  \alpha Q^2 - \beta Q & -\eta & 0               & \cdots & \cdots            & \cdots           & 0 \\
  -\eta & \alpha (Q-1)^2 - \beta (Q-1) & -\eta & 0       & \ddots            & \ddots           & \vdots \\
  0              &    \ddots           & \ddots         & \ddots & \ddots           &   \ddots               & \vdots \\
    \vdots              &    \ddots           & \ddots         & \ddots & \ddots           &   \ddots               & \vdots \\
      \vdots              &    \ddots           & \ddots         & \ddots & \ddots           &   \ddots               & 0 \\
                    \vdots& \ddots         & \ddots          & 0       & -\eta & \alpha (Q-1)^2 - \beta (Q-1)   & -\eta\\
 0                   & \cdots         & \cdots         & \cdots  & 0                & -\eta & \alpha Q^2 - \beta Q
\end{pmatrix}
\]
\end{Proposition}

In addition to this theoretical result, we can consider an approximation similar to the approximation used for the initial model with no drift. We then obtain the following approximations for the optimal quotes and the bid-ask spread:

$$\delta^{b*}_\infty(q) \simeq  \frac 1\gamma \ln\left(1+\frac \gamma k\right) + \left[-\frac{\mu}{\gamma\sigma^2} + \frac{2q+1}{2}\right] \sqrt{\frac{\sigma^2\gamma}{2kA}\left(1+\frac{\gamma}{k}\right)^{1+\frac{k}{\gamma}}} $$

$$\delta^{a*}_\infty(q) \simeq  \frac 1\gamma \ln\left(1+\frac \gamma k\right)  + \left[\frac{\mu}{\gamma\sigma^2} - \frac{2q-1}{2}\right] \sqrt{\frac{\sigma^2\gamma}{2kA}\left(1+\frac{\gamma}{k}\right)^{1+\frac{k}{\gamma}}} $$

$$\psi^*_\infty(q) \simeq\frac 2\gamma \ln\left(1+\frac \gamma k\right)+\sqrt{\frac{\sigma^2\gamma}{2kA}\left(1+\frac{\gamma}{k}\right)^{1+\frac{k}{\gamma}}}$$

\subsection{The case of market impact}

Another extension of the model consists in introducing market impact. The simplest way to proceed is to consider the following dynamics for the price:
 $$dS_t = \sigma dW_t + \xi dN^a_t - \xi dN^b_t, \qquad \xi > 0$$
When a limit order on the bid side is filled, the reference price decreases. On the contrary, when a limit order on the ask side is filled, the reference price increases. This is in line with the classical modeling of market impact for market orders, $\xi$ being a constant since the limit orders posted by the market maker are all supposed to be of the same size.\\
Adverse selection is another way to interpret the interaction we consider between the price process and the point processes modeling execution: trades on the bid side are often followed by a price decrease and, conversely, trades on the ask side are often followed by a price increase.\\

In this modified framework, the problem can be solved using a change of variables that is slightly more involved than the one presented above but the method is exactly the same and we have the following proposition (the proof is not repeated):

\begin{Proposition}[Solution with market impact]
Let us consider a family of functions $(v_q)_{|q|\le Q}$ solution of the linear system of ODEs that follows:
$$\forall q\in \lbrace -Q+1, \ldots, Q-1 \rbrace, \dot{v}_q(t) = \alpha q^2 v_q(t) - \eta e^{-\frac k2 \xi}  \left( v_{q-1}(t) + v_{q+1}(t) \right)$$
$$ \dot{v}_Q(t) = \alpha Q^2 v_Q(t) - \eta e^{-\frac k2 \xi} v_{Q-1}(t)$$
$$\dot{v}_{-Q}(t) = \alpha Q^2 v_{-Q}(t) - \eta e^{-\frac k2 \xi} v_{-Q+1}(t)$$
with $\forall q \in \lbrace -Q, \ldots, Q \rbrace,  v_q(T) = \exp(-\frac 12 k \xi q^2)$, where $\alpha = \frac k2 \gamma \sigma^2$ and $\eta = A(1+\frac \gamma k)^{-(1+\frac k\gamma)}$.\\

Then, $u(t,x,q,s) = -\exp(-\gamma(x+qs + \frac 12 \xi q^2)){v_q(t)}^{-\frac \gamma k}$ is the value function of the control problem.\\

The optimal quotes are given by:
$$s - s^{b*}(t,q,s) = \delta^{b*}(t,q) = \frac{1}{k}\ln\left(\frac{v_{q}(t)}{v_{q+1}(t)}\right) + \frac \xi2 + \frac 1\gamma \ln\left(1+\frac \gamma k\right)$$
$$s^{a*}(t,q,s) - s = \delta^{a*}(t,q) = \frac{1}{k}\ln\left(\frac{v_{q}(t)}{v_{q-1}(t)}\right) + \frac \xi2 +  \frac 1\gamma \ln\left(1+\frac \gamma k\right)$$
and the resulting bid-ask spread of the market maker is :
$$\psi^*(t,q) = -\frac{1}{k}\ln\left(\frac{v_{q+1}(t)v_{q-1}(t)}{v_q(t)^2}\right) + \xi + \frac 2\gamma \ln\left(1+\frac \gamma k\right)$$
Moreover,

$$\lim_{T\to+\infty} \delta^{b*}(0,q) = \frac 1\gamma \ln\left(1+\frac \gamma k\right) +\frac \xi 2 + \frac 1k \ln\left(\frac{f^0_{q}}{f^0_{q+1}}\right)$$
$$\lim_{T\to+\infty} \delta^{a*}(0,q) = \frac 1\gamma \ln\left(1+\frac \gamma k\right) + \frac \xi 2 + \frac 1k \ln\left(\frac{f^0_{q}}{f^0_{q-1}}\right)$$
$$\lim_{T\to+\infty} \psi^*(0,q) = -\frac{1}{k}\ln\left(\frac{f^0_{q+1}f^0_{q-1}}{{f^0_q}^2}\right) + \xi + \frac 2\gamma \ln\left(1+\frac \gamma k\right)$$
where $f^0$ is an eigenvector corresponding to the smallest eigenvalue of:
\[
 \begin{pmatrix}
  \alpha Q^2 & -\eta e^{-\frac k2 \xi} & 0               & \cdots & \cdots            & \cdots           & 0 \\
  -\eta e^{-\frac k2 \xi} & \alpha (Q-1)^2 & -\eta e^{-\frac k2 \xi} & 0       & \ddots            & \ddots           & \vdots \\
  0              &    \ddots           & \ddots         & \ddots & \ddots           &   \ddots               & \vdots \\
    \vdots              &    \ddots           & \ddots         & \ddots & \ddots           &   \ddots               & \vdots \\
      \vdots              &    \ddots           & \ddots         & \ddots & \ddots           &   \ddots               & 0 \\
                    \vdots& \ddots         & \ddots          & 0       & -\eta e^{-\frac k2 \xi} & \alpha (Q-1)^2   & -\eta e^{-\frac k2 \xi}\\
 0                   & \cdots         & \cdots         & \cdots  & 0                & -\eta e^{-\frac k2 \xi} & \alpha Q^2
\end{pmatrix}
\]
\end{Proposition}

In addition to this theoretical result, we can consider an approximation similar to the approximation used for the initial model. We then obtain the following approximations for the optimal quotes and the bid-ask spread:

$$\delta^{b*}_\infty(q) \simeq  \frac 1\gamma \ln\left(1+\frac \gamma k\right) + \frac \xi 2 + \frac{2q+1}{2} e^{\frac k 4 \xi}\sqrt{\frac{\sigma^2\gamma}{2kA}\left(1+\frac{\gamma}{k}\right)^{1+\frac{k}{\gamma}}} $$

$$\delta^{a*}_\infty(q) \simeq  \frac 1\gamma \ln\left(1+\frac \gamma k\right) + \frac \xi 2 - \frac{2q-1}{2} e^{\frac k 4 \xi} \sqrt{\frac{\sigma^2\gamma}{2kA}\left(1+\frac{\gamma}{k}\right)^{1+\frac{k}{\gamma}}} $$

$$\psi^*_\infty(q) \simeq\frac 2\gamma \ln\left(1+\frac \gamma k\right)+\xi +e^{\frac k 4 \xi}\sqrt{\frac{\sigma^2\gamma}{2kA}\left(1+\frac{\gamma}{k}\right)^{1+\frac{k}{\gamma}}}$$

\section{Comparative statics}

We argued in section 4 that the value of the optimal quotes was almost independent of $t$ for $t$ sufficiently far from the terminal time $T$ and we characterized, using spectral arguments, the asymptotic value of the optimal quotes. In order to obtain closed-form formulae, we also provided approximations for the asymptotic value of the optimal quotes. Although these closed-form formulae are only approximations, they provide a rather good intuition about the influence of the different parameters on the optimal quotes.

\subsection{Dependence on $\sigma^2$}

The dependence of optimal quotes on $\sigma^2$ depends on the sign of the inventory. More precisely, we observe numerically, in accordance with the approximations, that:

$$
\left\{
  \begin{array}{ll}
    \frac{\partial\delta_{\infty}^{b*}}{\partial{\sigma^2}} < 0,\quad  \frac{\partial\delta_{\infty}^{a*}}{\partial{\sigma^2}} > 0,  & \hbox{if $q<0$} \\
    \frac{\partial\delta_{\infty}^{b*}}{\partial{\sigma^2}} > 0,\quad \frac{\partial\delta_{\infty}^{a*}}{\partial{\sigma^2}} > 0, & \hbox{if $q=0$} \\
    \frac{\partial\delta_{\infty}^{b*}}{\partial{\sigma^2}} > 0,\quad \frac{\partial\delta_{\infty}^{a*}}{\partial{\sigma^2}} < 0, & \hbox{if $q>0$}
  \end{array}
\right.
$$

For the bid-ask spread, we obtain:
$$\frac{\partial\psi^*_{\infty}}{\partial{\sigma^2}} > 0$$

The rationale behind this is that an increase in $\sigma^2$ increases inventory risk. Hence, to reduce this risk, a market maker that has a long position will try to reduce his exposure and hence ask less for his stocks (to get rid of some of them) and accept to buy at a lower price (to avoid buying new stocks). Similarly, a market maker with a short position tries to buy stocks, and hence increases its bid quote, while avoiding short selling new stocks, and he increases its ask quote to that purpose. Overall, due to the increase in price risk, the bid-ask spread widens as it is well instanced in the case of a market maker with a flat position (this one wants indeed to earn more per trade to compensate the increase in inventory risk).

\subsection{Dependence on $\mu$}

The dependence of optimal quotes on the drift $\mu$ is straightforward and corresponds to the intuition. If the agent expect the price to increase (resp. decrease) he will post orders with higher (resp. lower) prices. Hence we have:

$$ \frac{\partial\delta_{\infty}^{b*}}{\partial{\mu}} < 0, \quad  \frac{\partial\delta_{\infty}^{a*}}{\partial{\mu}} > 0$$

\subsection{Dependence on $A$}

Because of the form of the system of equations that defines $v$, the dependence on $A$ must be the exact opposite of the dependence on $\sigma^2$:

$$
\left\{
  \begin{array}{ll}
    \frac{\partial\delta_{\infty}^{b*}}{\partial{A}} > 0,\quad \frac{\partial\delta_{\infty}^{a*}}{\partial{A}} < 0,  & \hbox{if $q<0$;} \\
    \frac{\partial\delta_{\infty}^{b*}}{\partial{A}} < 0,\quad \frac{\partial\delta_{\infty}^{a*}}{\partial{A}} < 0, & \hbox{if $q=0$} \\
    \frac{\partial\delta_{\infty}^{b*}}{\partial{A}} < 0,\quad \frac{\partial\delta_{\infty}^{a*}}{\partial{A}} > 0, & \hbox{if $q>0$}
  \end{array}
\right.
$$

For the bid-ask spread, we obtain:
$$\frac{\partial\psi^*_{\infty}}{\partial{A}} < 0$$

The rationale behind these results is that an increase in $A$ reduces the inventory risk. An increase in $A$ indeed increases the frequency of trades and hence reduces the risk of being stuck with a large inventory (in absolute value). For this reason, an increase in $A$ should have the same effect as a decrease in $\sigma^2$.

\subsection{Dependence on $\gamma$}

Using the closed-form approximations, we see that the dependence on $\gamma$ is ambiguous. The market maker faces indeed two different risks that contribute to inventory risk: (i) trades occur at random times and (ii) the reference price is stochastic. But if risk aversion increases, the market maker will mitigate the two risks: (i) he may set his quotes closer to one another to reduce the randomness in execution and (ii) he may widen his spread to reduce price risk. The tension between these two roles played by $\gamma$ explains the different behaviors we may observe, as on Figure \ref{imggamma1} and Figure \ref{imggamma2} for the bid-ask spread resulting from the asymptotic optimal quotes:

\begin{figure}[h]
\center
  % Requires \usepackage{graphicx}
  \includegraphics[width=0.85\textwidth]{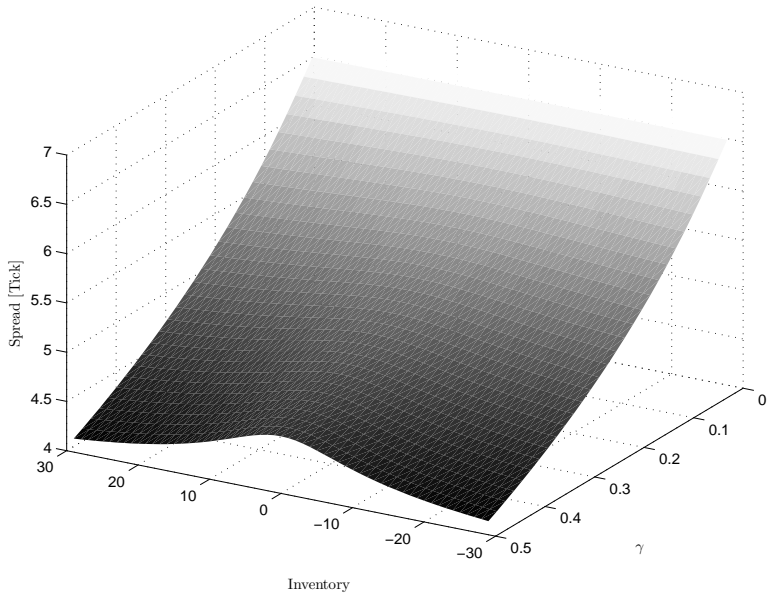}
  \caption{Bid-ask spread resulting from the asymptotic optimal quotes for different inventories and different values for the risk aversion parameter $\gamma$. $\sigma = 0.3\quad \mathrm{Tick}\cdot\mathrm{s}^{-1/2}$, $A = 0.9\quad\mathrm{s}^{-1}$, $k = 0.3\quad\mathrm{Tick}^{-1}$, $T = 600\quad s$.}
  \label{imggamma1}
\end{figure}

\begin{figure}[h]
\center
  % Requires \usepackage{graphicx}
  \includegraphics[width=0.85\textwidth]{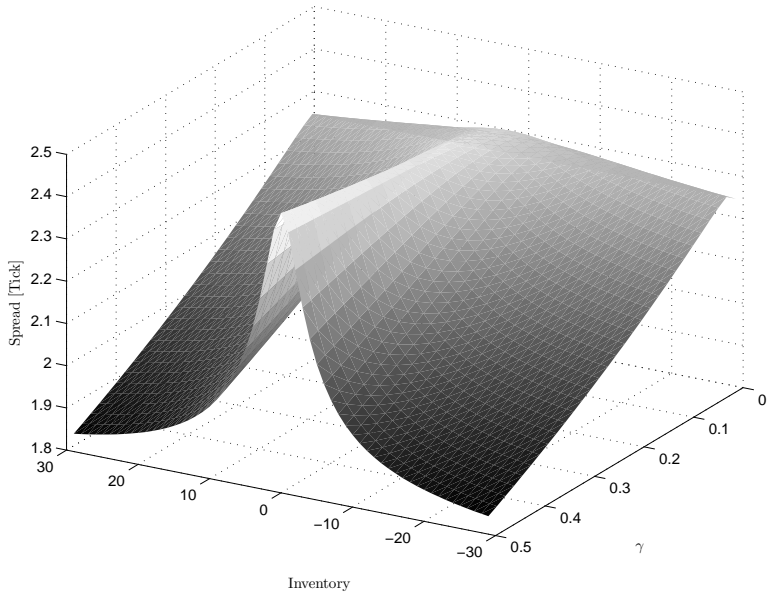}
  \caption{Bid-ask spread resulting from the asymptotic optimal quotes for different inventories and different values for the risk aversion parameter $\gamma$. $\sigma = 0.6\quad \mathrm{Tick}\cdot\mathrm{s}^{-1/2}$, $A = 0.9\quad\mathrm{s}^{-1}$, $k = 0.9\quad\mathrm{Tick}^{-1}$, $T = 600\quad s$.}
  \label{imggamma2}
\end{figure}

\subsection{Dependence on $k$}

From the closed-form approximations, we expect $\delta_{\infty}^{b*}$ to be decreasing in $k$ for $q$ greater than some negative threshold. Below this threshold, we expect it to be increasing. Similarly we expect $\delta_{\infty}^{a*}$ to be decreasing in $k$ for $q$ smaller than some positive threshold. Above this threshold we expect it to be increasing.\\

Eventually, as far as the bid-ask spread is concerned, the closed-form approximations indicate that the resulting bid-ask spread should be a decreasing function of $k$.
$$\frac{\partial\psi^*_{\infty}}{\partial{k}} < 0$$
In fact several effects are in interaction. On one hand, there is a ``no-volatility'' effect that is completely orthogonal to any reasoning on the inventory risk: when $k$ increases, in a situation where $\delta^b$ and $\delta^a$ are positive, trades occur closer to the reference price $S_t$. For this reason, and in absence of inventory risk, the optimal bid-ask spread has to shrink. However, an increase in $k$ also affects the inventory risk since it decreases the probability to be executed (for $\delta^b, \delta^a > 0$). Hence, an increase in $k$ is also, in some aspects, similar to a decrease in $A$. These two effects explain the expected behavior.\\

\begin{figure}[h]
\center
  % Requires \usepackage{graphicx}
  \includegraphics[width=0.85\textwidth]{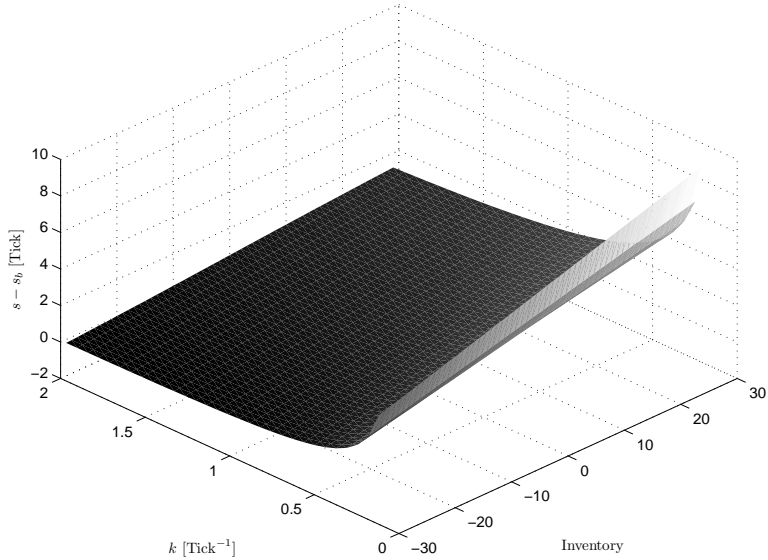}
  \caption{Asymptotic optimal bid quotes for different inventories and different values of $k$. $\sigma = 0.3\quad \mathrm{Tick}\cdot\mathrm{s}^{-1/2}$, $A = 0.9\quad\mathrm{s}^{-1}$, $\gamma = 0.01\quad \mathrm{Tick}^{-1}$, $T = 600\quad s$.}
  \label{imgk}
\end{figure}

Numerically, we observed that the ``no-volatility'' effect dominated for the values of the inventory under consideration (see Figure \ref{imgk} for the case of the bid quote\footnote{The case of the ask quote is obviously similar.}).

\subsection{Dependence on the market impact $\xi$}

The market impact introduced in 5.2 has two effects on the optimal quotes. In the absence of price risk, given the functional form of the execution intensities, the direct effect of $\xi$ is approximately to add $\frac \xi 2$ the each optimal quote: the market maker approximately maintains his profit per round trip on the market but the probability of occurrence of a trade is reduced. This adverse selection effect has a side-effect linked to inventory risk: since adverse selection gives the market maker an incentive to post orders deeper in the book, it increases the risk of being stuck with a large inventory for a market maker holding such an inventory. As a consequence, for a trader holding a positive (resp. negative) inventory, there is a second effect inciting to buy and sell at lower (resp. higher) prices. These two effects are clearly highlighted by the closed-form approximations exhibited in the previous section:

$$\delta^{b*}_\infty(q) \simeq  \frac 1\gamma \ln\left(1+\frac \gamma k\right) + \underbrace{\frac \xi 2}_{adverse\; selection} + \frac{2q+1}{2} \underbrace{e^{\frac k 4 \xi}}_{side-effect}\sqrt{\frac{\sigma^2\gamma}{2kA}\left(1+\frac{\gamma}{k}\right)^{1+\frac{k}{\gamma}}} $$

$$\delta^{a*}_\infty(q) \simeq  \frac 1\gamma \ln\left(1+\frac \gamma k\right) + \underbrace{\frac \xi 2}_{adverse\; selection} - \frac{2q-1}{2} \underbrace{e^{\frac k 4 \xi}}_{side-effect} \sqrt{\frac{\sigma^2\gamma}{2kA}\left(1+\frac{\gamma}{k}\right)^{1+\frac{k}{\gamma}}} $$

\section{Backtests}

Before using the above model on historical data, we need to discuss some features of the model that need to be adapted before any backtest is possible.\\

First of all, the model is continuous in both time and space while the real control problem is intrinsically discrete in space, because of the tick size, and in time, because orders have a certain priority and changing position too often reduces the actual chance to be reached by a market order. Hence, the model has to be reinterpreted in a discrete way. In terms of prices, quotes must not be between two ticks and we decided to round the optimal quotes to the nearest tick. In terms of time, an order of size ATS\footnote{ATS is the average trade size.} is sent to the market and is not canceled nor modified for a given period of time $\Delta t$, unless a trade occurs and, though perhaps partially, fills the order. Now, when a trade occurs and changes the inventory or when an order stayed in the order book for longer than $\Delta t$, then the optimal quote is updated and, if necessary, a new order is inserted.\\

Concerning the parameters, $\sigma$, $A$ and $k$ can be calibrated on trade-by-trade limit order book data while $\gamma$ has to be chosen. However, it is well known by practitioners that $A$ and $k$ have to depend at least on the actual market bid-ask spread. Since we do not explicitly take into account the underlying market, there is no market bid-ask spread in the model. For the backtest example we present below, $A$ and $k$ have been chosen independent of the spread but, in practice, the value of $A$ and $k$ are function of the market bid-ask spread.
As far as $\gamma$ is concerned, we decided in our backtests to assign $\gamma$ an arbitrary value for which the inventory stayed between -10 and 10 during the day we considered (the unit being the ATS).\\

Turning to the backtests, they were carried out with trade-by-trade data and we assumed that our orders were entirely filled when a trade occurred at or above the ask price quoted by the agent. Our goal here is just to exemplify the use of the model\footnote{We, voluntarily, do not give full details about the algorithm based on the model.} and we considered the case of the French stock France Telecom on March $15^{th}$ 2012.\\

\begin{figure}[!h]
\center
  \includegraphics[width=0.85\textwidth]{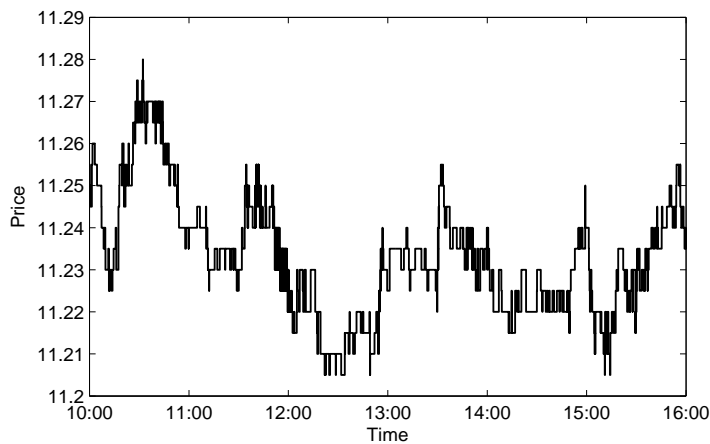}
  \caption{Price of the stock France Telecom on 15/03/2012, from 10:00 to 16:00.}
  \label{price}
\end{figure}

\begin{figure}[!h]
\center
  \includegraphics[width=0.85\textwidth]{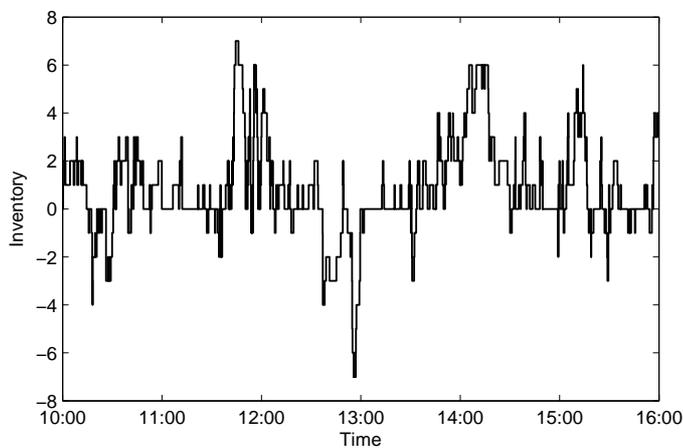}
  \caption{Inventory (in ATS) when the strategy is used on France Telecom (15/03/2012) from 10:00 to 16:00.}
  \label{inventory}
\end{figure}

\begin{figure}[!h]
\center
  \includegraphics[width=0.85\textwidth]{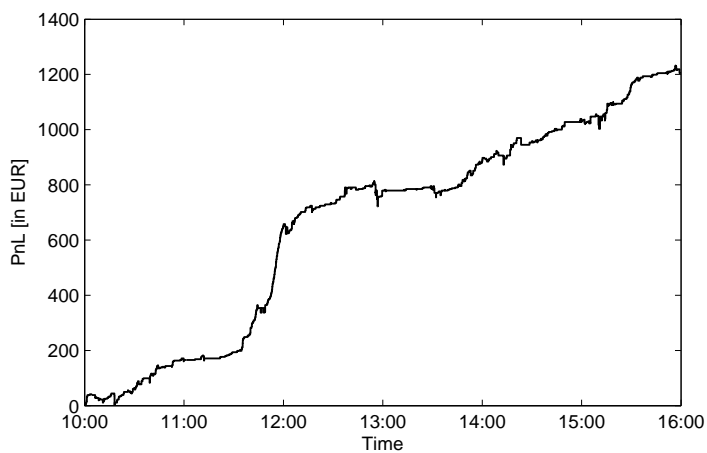}
  \caption{P\&L when the strategy is used on France Telecom (15/03/2012) from 10:00 to 16:00.}
  \label{pnl}
\end{figure}

We first plot the price of the stock France Telecom on March $15^{th}$ 2012 on Figure \ref{price}, the evolution of the inventory\footnote{The ATS is $1105$ for the day we considered.} on Figure \ref{inventory} and the associated P\&L (the stocks in the portfolio are evaluated at mid-price) on Figure \ref{pnl}.
\vspace{0.8cm}
\vspace{0.5cm}
\vspace{0.8cm}

\begin{figure}[!h]
\center
  \includegraphics[width=0.85\textwidth]{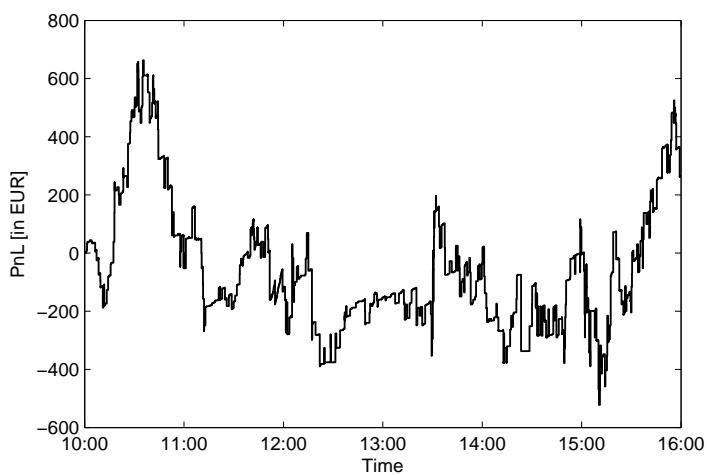}
  \caption{P\&L of a naive market maker on France Telecom (15/03/2012) from 10:00 to 16:00.}
  \label{nai}
\end{figure}

This P\&L can be compared to the P\&L of a naive trader (Figure \ref{nai}) who only posts orders at the first limit of the book on each side, whenever he is asked to post orders -- that is when one of his orders has been executed or after a period of time $\Delta t$ with no execution.

\begin{figure}[h!]
\center
  \includegraphics[width=0.85\textwidth]{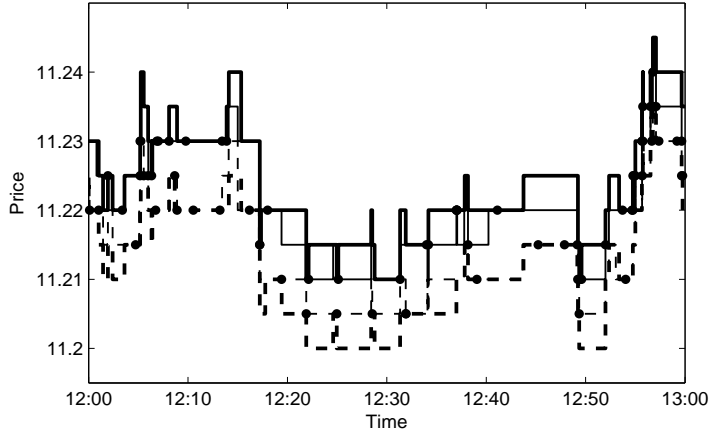}
    \caption{Details for the quotes and trades when the strategy is used on France Telecom (15/03/2012). Thin lines represent the market while bold lines represent the quotes of the market maker. Dotted lines are associated to the bid side while plain lines are associated to the ask side. Black points represent trades in which the market maker is involved.}
  \label{execution}
\end{figure}

Now, to better understand the details of the strategy, we focused on a subperiod of 1 hour and we plotted the state of the market along with the quotes of the market maker (Figure \ref{execution}). Trades occurrences involving the market maker are signalled by a dot.

\section*{Conclusion}

In this paper we present a model for the optimal quotes of a market maker. Starting from a model in line with Avellaneda and Stoikov \cite{avellaneda2008high} and rooted to Ho and Stoll \cite{ho1981optimal}, we introduce a change of variables that allows to transform the HJB equation into a system of linear ordinary differential equations. This transformation allows to find the optimal quotes and to characterize their asymptotic behavior. Closed-form approximations are also obtained using spectral analysis.\\

The change of variables introduced in this paper can also be used to solve the initial equations of Avellaneda and Stoikov \cite{avellaneda2008high} and we provide a complete mathematical proof in \cite{gueant2012PDE}. However, in the absence of inventory limits, no proof of optimality is available for the quotes claimed to be optimal in \cite{avellaneda2008high} and their admissibility appears to be an open problem.\\

An important topic for future research consists in generalizing the model to any intensity function. This is particularly important because the exponential form of the intensity is only suited to liquid stocks with a small bid-ask spread. Another important topic consists in introducing ``passive market impact'' (\emph{i.e.} the perturbations of the price formation process by liquidity provision). This is a real modeling challenge since no quantitative model for this type of impact has been proposed in the literature.

\begin{acknowledgements}
The authors wish to acknowledge the helpful conversations with Yves Achdou (Universit\'e Paris-Diderot), Vincent Fardeau (London School of Economics), Thierry Foucault (HEC), Jean-Michel Lasry (Universit\'e Paris-Dauphine), Antoine Lemenant (Universit\'e Paris-Diderot), Pierre-Louis Lions (Coll\`ege de France), Albert Menkveld (VU University Amsterdam), Vincent Millot (Universit\'e Paris-Diderot) and Nizar Touzi (Ecole Polytechnique). The authors also would like to thank two anonymous referees for their suggestions.
\end{acknowledgements}

\section*{Appendix: Proofs of the results}

\noindent\textbf{Proof of Proposition 1, Proposition 2 and Theorem 1:}\\

Let us consider a family $(v_q)_{|q|\le Q}$ of  positive functions solution of the system of ODEs introduced in Proposition 1 and let us define $u(t,x,q,s) = -\exp\left(-\gamma (x+qs)\right){v_q(t)}^{-\frac \gamma k}$.\\

Then:
$$\partial_t u + \frac 12 \sigma^2 \partial^2_{ss} u =-\frac \gamma k \frac{\dot{v}_q(t)}{v_q(t)} u + \frac{\gamma^2\sigma^2}{2} q^2 u$$

Now, concerning the hamiltonian parts, we have for the bid part ($q \neq Q$):
$$\sup_{\delta^b} \lambda^b(\delta^b) \left[u(t,x-s + \delta^b,q+1,s) - u(t,x,q,s) \right]$$
$$=\sup_{\delta^b} A e^{-k\delta^b} u(t,x,q,s) \left[\exp(- \gamma \delta^b)\left(\frac{v_{q+1}(t)}{v_q(t)}\right)^{-\frac{\gamma}{k}} - 1 \right] $$

The first order condition of this problem corresponds to a maximum (because $u$ is negative) and writes:
$$(k+\gamma) \exp(-\gamma \delta^{b*})\left(\frac{v_{q+1}(t)}{v_q(t)}\right)^{-\frac{\gamma}{k}} = k$$

Hence:

$$\delta^{b*} = \frac 1k \ln\left(\frac{v_{q}(t)}{v_{q+1}(t)}\right) + \frac 1\gamma \ln\left(1+\frac \gamma k\right)$$

and

$$\sup_{\delta^{b}} \lambda^b(\delta^{b}) \left[u(t,x-s+\delta^{b},q+1,s) - u(t,x,q,s) \right]$$$$ = -\frac{\gamma}{k+\gamma}A\exp(-k\delta^{b*})u(t,x,q,s)$$$$=-\frac{\gamma A}{k+\gamma} \left(1+\frac \gamma k\right)^{-\frac k \gamma} \frac{v_{q+1}(t)}{v_q(t)}  u(t,x,q,s)$$

Similarly, the maximizer for the ask part (for $q \neq -Q$) is:

$$\delta^{a*} = \frac 1k \ln\left(\frac{v_{q}(t)}{v_{q-1}(t)}\right) + \frac 1\gamma \ln\left(1+\frac \gamma k\right)$$

and

$$\sup_{\delta^{a}} \lambda^a(\delta^{a}) \left[u(t,x+s+\delta^{a},q-1,s) - u(t,x,q,s) \right]$$$$ = -\frac{\gamma}{k+\gamma}A\exp(-k\delta^{a*})u(t,x,q,s)$$$$=-\frac{\gamma A}{k+\gamma} \left(1+\frac \gamma k\right)^{-\frac k \gamma} \frac{v_{q-1}(t)}{v_q(t)}  u(t,x,q,s)$$

Hence, putting the terms altogether we get for $|q| < Q$:
$$\partial_t u(t,x,q,s) + \frac 12 \sigma^2 \partial^2_{ss} u(t,x,q,s)$$$$ + \sup_{\delta^{b}} \lambda^b(\delta^{b}) \left[u(t,x-s+\delta^{b},q+1,s) - u(t,x,q,s) \right]$$$$ + \sup_{\delta^{a}} \lambda^a(\delta^{a}) \left[u(t,x+s+\delta^{a},q-1,s) - u(t,x,q,s) \right]$$
$$=-\frac \gamma k \frac{\dot{v}_q(t)}{v_q(t)} u + \frac{\gamma^2\sigma^2}{2} q^2 u -\frac{\gamma A}{k+\gamma} \left(1+\frac \gamma k\right)^{\frac k \gamma} \left[\frac{v_{q+1}(t)}{v_q(t)} + \frac{v_{q-1}(t)}{v_q(t)}\right]  u $$
$$=-\frac \gamma k  \frac{u}{v_q(t)} \left[\dot{v}_q(t) - \frac{k\gamma\sigma^2}{2} q^2 v_q(t) + A \left(1+\frac \gamma k\right)^{-\left(1+\frac k \gamma\right)} (v_{q+1}(t)+v_{q-1}(t))  \right]=0$$

For $q=-Q$ we have:

$$\partial_t u(t,x,q,s) + \frac 12 \sigma^2 \partial^2_{ss} u(t,x,q,s)$$$$ + \sup_{\delta^{b}} \lambda^b(\delta^{b}) \left[u(t,x-s+\delta^{b},q+1,s) - u(t,x,q,s) \right]$$
$$=-\frac \gamma k \frac{\dot{v}_q(t)}{v_q(t)} u + \frac{\gamma^2\sigma^2}{2} q^2 u -\frac{\gamma A}{k+\gamma} \left(1+\frac \gamma k\right)^{\frac k \gamma} \frac{v_{q+1}(t)}{v_q(t)} u $$
$$=-\frac \gamma k  \frac{u}{v_q(t)} \left[\dot{v}_q(t) - \frac{k\gamma\sigma^2}{2} q^2 v_q(t) + A \left(1+\frac \gamma k\right)^{-\left(1+\frac k \gamma\right)} v_{q+1}(t)  \right]=0$$

Similarly, for $q=Q$ we have:

$$\partial_t u(t,x,q,s) + \frac 12 \sigma^2 \partial^2_{ss} u(t,x,q,s)$$$$ + \sup_{\delta^{a}} \lambda^a(\delta^{a}) \left[u(t,x-s+\delta^{a},q+1,s) - u(t,x,q,s) \right]$$
$$=-\frac \gamma k \frac{\dot{v}_q(t)}{v_q(t)} u + \frac{\gamma^2\sigma^2}{2} q^2 u -\frac{\gamma A}{k+\gamma} \left(1+\frac \gamma k\right)^{\frac k \gamma} \frac{v_{q-1}(t)}{v_q(t)} u $$
$$=-\frac \gamma k  \frac{u}{v_q(t)} \left[\dot{v}_q(t) - \frac{k\gamma\sigma^2}{2} q^2 v_q(t) + A \left(1+\frac \gamma k\right)^{-\left(1+\frac k \gamma\right)} v_{q-1}(t)  \right]=0$$

Now, noticing that the terminal condition for $v_q$ is consistent with the terminal condition for $u$, we get that $u$ verifies (HJB) and this proves Proposition 1.\\

The positivity of the functions $(v_q)_{|q|\le Q}$ was essential in the definition of $u$. Hence we need to prove that the solution to the above linear system of ordinary differential equations, namely $v(t) = \exp(-M(T-t)) \times (1, \dots, 1)'$ (where $M$ is given in Proposition 2), defines a family $(v_q)_{|q|\le Q}$ of  positive functions.\\

In fact, we are going to prove that: $$\forall t \in [0,T], \forall q \in \lbrace -Q, \ldots, Q \rbrace,\quad  v_q(t) \ge e^{-(\alpha Q^2 - \eta)(T-t)}$$

If this was not true then there would exist $\epsilon > 0$ such that:

$$\min_{t \in [0,T], |q| \le Q} e^{-2\eta (T-t)} \left(v_q(t) - e^{-(\alpha Q^2 - \eta)(T-t)} \right) + \epsilon (T-t) < 0$$

But this minimum is achieved at some point $(t^*,q^*)$ with $t^* < T$ and hence:

$$\frac{d\ }{dt} \left.e^{-2\eta (T-t)} \left(v_{q^*}(t) - e^{-(\alpha Q^2 - \eta)(T-t)} \right)\right|_{t=t^*} \ge \epsilon$$

This gives:

 $$2\eta e^{-2\eta (T-t^*)} \left(v_{q^*}(t^*) - e^{-(\alpha Q^2 - \eta)(T-t^*)} \right)$$$$ + e^{-2\eta (T-t^*)} \left(v'_{q^*}(t^*) - (\alpha Q^2 - \eta)e^{-(\alpha Q^2 - \eta)(T-t^*)} \right) \ge \epsilon$$
 Hence:
 $$ 2\eta v_{q^*}(t^*) + v'_{q^*}(t^*) - (\eta + \alpha Q^2) e^{-(\alpha Q^2 - \eta)(T-t^*)} \ge \epsilon e^{2\eta (T-t^*)}$$

 Now, if $|{q^*}|< Q$, this gives:
 $$ \alpha {q^*}^2 v_{q^*}(t^*) -\eta ( v_{{q^*}+1}(t^*) - 2 v_{q^*}(t^*) + v_{{q^*}-1}(t^*))$$$$ - (\eta + \alpha Q^2) e^{-(\alpha Q^2 - \eta)(T-t^*)} \ge \epsilon e^{2\eta (T-t^*)}$$
 Thus:
 $$ \alpha {q^*}^2 \left(v_{q^*}(t^*) - e^{-(\alpha Q^2 - \eta)(T-t^*)}\right) -\eta ( v_{{q^*}+1}(t^*) - 2 v_{q^*}(t^*) + v_{{q^*}-1}(t^*))$$$$ - (\eta + \alpha (Q^2 - {q^*}^2)) e^{-(\alpha Q^2 - \eta)(T-t^*)} \ge \epsilon e^{2\eta (T-t^*)}$$

 All the terms on the left hand side are nonpositive by definition of $(t^*,q^*)$ and this gives a contradiction.\\

 If $q^*=Q$, we have:

 $$ (\alpha {Q}^2 + \eta) v_{Q}(t^*) -\eta (v_{Q-1}(t^*) - v_{Q}(t^*))$$$$ - (\eta + \alpha Q^2) e^{-(\alpha Q^2 - \eta)(T-t^*)} \ge \epsilon e^{2\eta (T-t^*)}$$
 Thus:
 $$ -\eta ( v_{Q-1}(t^*) -  v_{Q}(t^*)) + (\eta + \alpha Q^2)\left( v_{Q}(t^*) -e^{-(\alpha Q^2 - \eta)(T-t^*)} \right) \ge \epsilon e^{2\eta (T-t^*)}$$

 All the terms on the left hand side are nonpositive by definition of $(t^*,q^*)=(t^*,Q)$ and this gives a contradiction.\\

 Similarly, if $q^*=-Q$, we have:

$$ (\alpha {Q}^2 + \eta) v_{-Q}(t^*) -\eta (v_{-Q+1}(t^*) - v_{Q}(t^*))$$$$ - (\eta + \alpha Q^2) e^{-(\alpha Q^2 - \eta)(T-t^*)} \ge \epsilon e^{2\eta (T-t^*)}$$
$$ -\eta ( v_{-Q+1}(t^*) -  v_{-Q}(t^*)) + (\eta + \alpha Q^2)\left( v_{-Q}(t^*) -e^{-(\alpha Q^2 - \eta)(T-t^*)} \right) \ge \epsilon e^{2\eta (T-t^*)}$$

All the terms on the left hand side are nonpositive by definition of $(t^*,q^*)=(t^*,-Q)$ and this gives a contradiction.\\

As a consequence, $v_q(t) \ge e^{-(\alpha Q^2 - \eta)(T-t)} > 0$ and this completes the proof of Proposition 2.\\

Combining the above results, we see that $u$, as defined in Theorem 1, is a solution of (HJB). Then, we are going to use a verification argument to prove that $u$ is the value function of the optimal control problem under consideration and prove subsequently that the optimal controls are as given in Theorem 1.\\

Let us consider processes $(\nu^b)$ and $(\nu^a) \in \mathcal{A}$. Let $t \in [0,T)$ and let us consider the following processes for $\tau \in [t,T]$:

$$dS^{t,s}_\tau = \sigma dW_\tau, \qquad S^{t,s}_t = s$$
$$dX^{t,x,\nu}_\tau = (S_\tau + \nu^a_\tau)  dN^a_\tau - (S_\tau - \nu^b_\tau)  dN^b_\tau , \qquad  X^{t,x,\nu}_t = x$$
$$dq^{t,q,\nu}_\tau = dN^b_\tau - dN^a_\tau, \qquad  q^{t,q,\nu}_t = q$$
where the point process $N^b$ has intensity $(\lambda^b_\tau)_\tau$ with $\lambda^b_\tau = A e^{-k\nu^b_\tau} 1_{q_{\tau-} < Q}$ and where the point process $N^a$ has intensity $(\lambda^a_\tau)_\tau$ with $\lambda^a_\tau = A e^{-k\nu^a_\tau} 1_{q_{\tau-} > -Q}$ \footnote{These intensities are bounded since $\nu^b$ and $\nu^a$ are bounded from below.}.\\

Now, since $u$ is smooth, let us write It\^o's formula for $u$, between $t$ and $t_n$ where $t_n = T \land \inf \lbrace \tau > t, |S_\tau - s| \ge n \mathrm{\; or\; } |N^a_\tau - N^a_t|\ge n \mathrm{\; or\; } |N^b_\tau - N^b_t|\ge n \rbrace$ ($n \in \mathbb{N}$):

$$u(t_n,X^{t,x,\nu}_{t_n-},q^{t,q,\nu}_{t_n-},S^{t,s}_{t_n}) = u(t,x,q,s)$$$$ + \int_t^{t_n} \left(\partial_\tau u(\tau,X^{t,x,\nu}_{\tau-},q^{t,q,\nu}_{\tau-},S^{t,s}_{\tau}) +\frac {\sigma^2}2 \partial^2_{ss} u(\tau,X^{t,x,\nu}_{\tau-},q^{t,q,\nu}_{\tau-},S^{t,s}_\tau)\right)d\tau$$
$$+ \int_t^{t_n}  \left(u(\tau,X^{t,x,\nu}_{\tau-}+S^{t,s}_\tau + \nu^a_\tau,q^{t,q,\nu}_{\tau-}-1,S^{t,s}_\tau) - u(\tau,X^{t,x,\nu}_{\tau-},q^{t,q,\nu}_{\tau-},S^{t,s}_\tau)\right) \lambda^a_\tau d\tau
$$
$$+ \int_t^{t_n}  \left(u(\tau,X^{t,x,\nu}_{\tau-}-S^{t,s}_\tau + \nu^b_\tau,q^{t,q,\nu}_{\tau-}+1,S^{t,s}_\tau) - u(\tau,X^{t,x,\nu}_{\tau-},q^{t,q,\nu}_{\tau-},S^{t,s}_\tau)\right) \lambda^b_\tau d\tau
$$
$$  +\int_t^{t_n} \sigma \partial_s u(\tau,X^{t,x,\nu}_{\tau-},q^{t,q,\nu}_{\tau-},S^{t,s}_\tau) dW_{\tau}$$
$$ + \int_t^{t_n}  \left(u(\tau,X^{t,x,\nu}_{\tau-}+S^{t,s}_\tau + \nu^a_\tau,q^{t,q,\nu}_{\tau-}-1,S^{t,s}_\tau) - u(\tau,X^{t,x,\nu}_{\tau-},q^{t,q,\nu}_{\tau-},S^{t,s}_\tau)\right) dM^a_\tau$$
$$ + \int_t^{t_n}  \left(u(\tau,X^{t,x,\nu}_{\tau-}-S^{t,s}_\tau + \nu^b_\tau,q^{t,q,\nu}_{\tau-}+1,S^{t,s}_\tau) - u(\tau,X^{t,x,\nu}_{\tau-},q^{t,q,\nu}_{\tau-},S^{t,s}_\tau)\right) dM^b_\tau$$
where $M^b$ and $M^a$ are the compensated processes associated respectively to $N^b$ and $N^a$ for the intensity processes $(\lambda^b_\tau)_\tau$ and $(\lambda^a_\tau)_\tau$.\\

Now, because each $v_q$ is continuous and positive on a compact set, it has a  positive lower bound and $v_{q_\tau}(\tau)^{-\frac{\gamma}{k}}$ is bounded along the trajectory, independently of the trajectory. Also, because $\nu^b$ and $\nu^a$ are bounded from below, and because of the definition of $t_n$, all the terms in the above stochastic integrals are bounded and, local martingales being in fact martingales, we have:

$$\mathbb{E}\left[u(t_n,X^{t,x,\nu}_{t_n-},q^{t,q,\nu}_{t_n-},S^{t,s}_{t_n})\right] = u(t,x,q,s)$$$$ + \mathbb{E}\left[\int_t^{t_n} \left(\partial_\tau u(\tau,X^{t,x,\nu}_{\tau-},q^{t,q,\nu}_{\tau-},S^{t,s}_{\tau}) +\frac {\sigma^2}2 \partial^2_{ss} u(\tau,X^{t,x,\nu}_{\tau-},q^{t,q,\nu}_{\tau-},S^{t,s}_\tau)\right)d\tau\right.$$
$$+ \int_t^{t_n}  \left(u(\tau,X^{t,x,\nu}_{\tau-}+S^{t,s}_\tau + \nu^a_\tau,q^{t,q,\nu}_{\tau-}-1,S^{t,s}_\tau) - u(\tau,X^{t,x,\nu}_{\tau-},q^{t,q,\nu}_{\tau-},S^{t,s}_\tau)\right) \lambda^a_\tau d\tau
$$
$$\left.+ \int_t^{t_n}  \left(u(\tau,X^{t,x,\nu}_{\tau-}-S^{t,s}_\tau + \nu^b_\tau,q^{t,q,\nu}_{\tau-}+1,S^{t,s}_\tau) - u(\tau,X^{t,x,\nu}_{\tau-},q^{t,q,\nu}_{\tau-},S^{t,s}_\tau)\right) \lambda^b_\tau d\tau\right]
$$

Using the fact that $u$ solves (HJB), we then have that $$\mathbb{E}\left[u(t_n,X^{t,x,\nu}_{t_n-},q^{t,q,\nu}_{t_n-},S^{t,s}_{t_n})\right] \le u(t,x,q,s)$$
with equality when the controls are taken equal the maximizers of the hamiltonians (these controls being in $\mathcal{A}$ because $v$ is bounded and has a  positive lower bounded).\\

Now, if we prove that $$\lim_{n \to \infty} \mathbb{E}\left[u(t_n,X^{t,x,\nu}_{t_n-},q^{t,q,\nu}_{t_n-},S^{t,s}_{t_n})\right] = \mathbb{E}\left[u(T,X^{t,x,\nu}_{T},q^{t,q,\nu}_{T},S^{t,s}_{T})\right]$$ we will have that for all controls in $\mathcal{A}$:

$$\mathbb{E}\left[-\exp\left(-\gamma (X^{t,x,\nu}_{T}+q^{t,q,\nu}_{T}S^{t,s}_{T}) \right)\right] = \mathbb{E}\left[u(T,X^{t,x,\nu}_{T},q^{t,q,\nu}_{T},S^{t,s}_{T})\right] \le u(t,x,q,s)$$ with equality for $\nu^b_t = \delta^{b*}(t,q_{t-})$ and $\nu^a_t = \delta^{a*}(t,q_{t-})$. Hence:

$$\sup_{(\nu_t^a)_t,(\nu_t^b)_t \in \mathcal{A}} \mathbb{E}\left[-\exp\left(-\gamma (X^{t,x,\nu}_{T}+q^{t,q,\nu}_{T}S^{t,s}_{T}) \right)\right] = u(t,x,q,s)$$$$ =
\mathbb{E}\left[-\exp\left(-\gamma (X^{t,x,\delta^*}_{T}+q^{t,q,\delta^*}_{T}S^{t,s}_{T}) \right)\right]$$
and this will give the result.\\

It remains to prove that $$\lim_{n \to \infty} \mathbb{E}\left[u(t_n,X^{t,x,\nu}_{t_n-},q^{t,q,\nu}_{t_n-},S^{t,s}_{t_n})\right] = \mathbb{E}\left[u(T,X^{t,x,\nu}_{T},q^{t,q,\nu}_{T},S^{t,s}_{T})\right]$$

First, we have, almost surely, that $u(t_n,X^{t,x,\nu}_{t_n-},q^{t,q,\nu}_{t_n-},S^{t,s}_{t_n})$ tends towards $u(T,X^{t,x,\nu}_{T-},q^{t,q,\nu}_{T-},S^{t,s}_{T})$. Then, in order to prove that the sequence is uniformly integrable we will bound it in $L^2$. However, because of the uniform lower bound on $v$ already used early, it is sufficient to bound $\exp(-\gamma(X^{t,x,\nu}_{t_n-}+q^{t,q,\nu}_{t_n-}S^{t,s}_{t_n}))$ in $L^2$.\\

But,

$$X^{t,x,\nu}_{t_n-}+q^{t,q,\nu}_{t_n-}S^{t,s}_{t_n} = \int_t^{t_n} \nu^a_\tau dN^a_\tau + \int_t^{t_n} \nu^b_\tau dN^b_\tau + \sigma \int_t^{t_n} q^{t,q,\nu}_{\tau} dW_\tau$$
$$ \ge - \|\nu^a_-\|_{\infty} N^a_T - \|\nu^b_-\|_{\infty} N^b_T + \sigma \int_t^{t_n} q^{t,q,\nu}_{\tau} dW_\tau$$

Hence

\begin{eqnarray*}
% \nonumber to remove numbering (before each equation)
  & & \mathbb{E}\left[\exp(-2\gamma(X^{t,x,\nu}_{t_n-}+q^{t,q,\nu}_{t_n-}S^{t,s}_{t_n}))\right]\\
   &\le & \mathbb{E}\left[\exp\left(2\gamma \|\nu^a_-\|_{\infty} N^a_T\right)\exp\left(2\gamma \|\nu^a_-\|_{\infty} N^b_T\right)\exp\left(-2\gamma \sigma \int_t^{t_n} q^{t,q,\nu}_{\tau} dW_\tau\right)\right]  \\
   &\le & \mathbb{E}\left[\exp\left(6\gamma \|\nu^a_-\|_{\infty} N^a_T\right)\right]^{\frac 13} \mathbb{E}\left[\exp\left(6\gamma \|\nu^b_-\|_{\infty} N^b_T\right)\right]^{\frac 13}\\
   &&\times\mathbb{E}\left[\exp\left(-6\gamma \sigma \int_t^{t_n} q^{t,q,\nu}_{\tau} dW_\tau\right)\right]^{\frac 13} \\
\end{eqnarray*}

Now, since the intensity of each point process is bounded, the point processes have a Laplace transform and the first two terms of the product are finite (and independent of $n$). Concerning the third term, because $|q^{t,q,\nu}_{\tau}|$ is bounded by $Q$, we know (for instance applying Girsanov's theorem) that:

$$\mathbb{E}\left[\exp\left(-6\gamma \sigma \int_t^{t_n} q^{t,q,\nu}_{\tau} dW_\tau\right)\right]^{\frac 13} \le \mathbb{E}\left[\exp\left(3\gamma^2 \sigma^2 (t_n -t) Q^2\right)\right]^{\frac 13}$$$$ \le \exp\left(\gamma^2 \sigma^2 Q^2 T\right) $$

Hence, the sequence is bounded in $L^2$, then uniformly integrable and we have:

$$\lim_{n \to \infty} \mathbb{E}\left[u(t_n,X^{t,x,\nu}_{t_n-},q^{t,q,\nu}_{t_n-},S^{t,s}_{t_n})\right] = \mathbb{E}\left[u(T,X^{t,x,\nu}_{T-},q^{t,q,\nu}_{T-},S^{t,s}_{T})\right]$$$$=\mathbb{E}\left[u(T,X^{t,x,\nu}_{T},q^{t,q,\nu}_{T},S^{t,s}_{T})\right]$$

We have proved that $u$ is the value function and that $\delta^{b*}$ and $\delta^{a*}$ are optimal controls.\\\qed

\noindent\textbf{Proof of Theorem 2:}\\

Let us first consider the matrix $M+2\eta I$. This matrix is a symmetric matrix and it is therefore diagonalizable. Its smallest eigenvalue $\lambda$ is characterized by:

$$\lambda = \underset{x \in \mathbb{R}^{2Q+1}\setminus \lbrace 0 \rbrace}{\inf} \frac{x'(M+2\eta I) x}{x'x}$$
and the associated eigenvectors $x \ne 0$ are characterized by:
$$\lambda =\frac{x'(M+2\eta I) x}{x'x}$$

It is straightforward to see that:

$$x'(M+2\eta I) x = \sum_{q=-Q}^Q \alpha q^2 {x_q}^2 + \eta \sum_{q=-Q}^{Q-1} (x_{q+1} - x_q)^2 + \eta {x_Q}^2 + \eta {x_{-Q}}^2$$

Hence, if $x$ is an eigenvector of $M+2\eta I$ associated to $\lambda$:

$$\lambda \le \frac{|x|'(M+2\eta I) |x|}{|x|'|x|}$$
$$ = \frac {1}{|x|'|x|}\left[\sum_{q=-Q}^Q \alpha q^2 {|x_q|}^2 + \eta \sum_{q=-Q}^{Q-1} (|x_{q+1}| - |x_q|)^2 + \eta {|x_Q|}^2 + \eta {|x_{-Q}|}^2\right]$$
$$\le \frac {1}{|x|'|x|}\left[\sum_{q=-Q}^Q \alpha q^2 {|x_q|}^2 + \eta \sum_{q=-Q}^{Q-1} (x_{q+1} - x_q)^2 + \eta {|x_Q|}^2 + \eta {|x_{-Q}|}^2\right] = \lambda$$

This proves that $|x|$ is also an eigenvector and that necessarily $x_{q+1}$ and $x_q$ are of the same sign (\emph{i.e.} $x_q x_{q+1} \ge 0$).\\

Now, let $x\ge 0$ be an eigenvector of $M+2\eta I$ associated to $\lambda$.\\

If for some $q$ with $|q| < Q$ we have $x_q = 0$ then:

$$0 = \lambda x_q = \alpha {q}^2 x_{q} - \eta (x_{q+1} - 2 x_{q} + x_{q-1}) = - \eta (x_{q+1} + x_{q-1}) \le 0$$

Hence, because $x \ge 0$, both $x_{q+1}$ and $x_{q-1}$ are equal to $0$. By immediate induction $x=0$ and this is a contradiction.\\

Now, if $x_Q = 0$, then $0 = \lambda x_Q = \alpha {Q}^2 x_{Q} - \eta ( - 2 x_{Q} + x_{Q-1}) = - \eta  x_{Q-1} \le 0$ and hence $x_{Q-1}=0$. Then, by the preceding reasoning we obtain a contradiction.\\

Similarly if $x_{-Q} = 0$, then $0 = \lambda x_{-Q} = \alpha {Q}^2 x_{-Q} - \eta ( x_{-Q+1}- 2 x_{-Q}) = - \eta  x_{-Q+1} \le 0$ and hence $x_{-Q+1}=0$. Then, as above, we obtain a contradiction.\\

This proves that any eigenvector $x\ge 0$ of $M+2\eta I$ associated to $\lambda$ verifies in fact $x>0$.\\

Now, if the eigenvalue $\lambda$ was not simple, there would exist two eigenvectors $x$ and $y$ of $M+2\eta I$ associated to $\lambda$ such that $|x|'y = 0$. Hence, $y$ must have positive coordinates and negative coordinates and since $y_{q} y _{q+1} \ge 0$, we know that there must exist $q$ such that $y_q = 0$. However, this contradicts our preceding point since $|y| \ge 0$ should also be an eigenvector of $M+2\eta I$ associated to $\lambda$ and it cannot have therefore coordinates equal to $0$.\\

As a conclusion, the eigenspace of $M+2\eta I$ associated to $\lambda$ is spanned by a vector $f^0 > 0$ and we scaled its $\mathbb{R}^{2Q+1}$-norm to $1$.

Now, because $M$ is a symmetric matrix, we can write $v(0) = \exp(-MT) \times (1, \ldots, 1)'$ as:

$$v_q(0) = \sum_{i=0}^{2Q} \exp(-\lambda^i T) \langle g^i, (1, \ldots, 1)'\rangle g_q^i, \qquad \forall q \in \lbrace -Q, \ldots, Q \rbrace$$ where $\lambda^0 \le \lambda^1 \le \ldots \le \lambda^{2Q}$ are the eigenvalues of $M$ (in increasing order and repeated if necessary) and $(g^i)_i$ an associated orthonormal basis of eigenvectors. Clearly, we can take $g^0 = f^0$. Then, both $f^0_q$ and $\langle f^0, (1, \ldots, 1)'\rangle$ are positive and hence different from zero. As a consequence:

$$v_q(0) \sim_{T \to +\infty} \exp(-\lambda^0 T) \langle f^0, (1, \ldots, 1)'\rangle f_q^0, \qquad \forall q \in \lbrace -Q, \ldots, Q \rbrace$$

Then, using the expressions for the optimal quotes, we get:

$$\lim_{T\to+\infty} \delta^{b*}(0,q) = \frac 1\gamma \ln\left(1+\frac \gamma k\right) + \frac 1k \ln\left(\frac{f^0_{q}}{f^0_{q+1}}\right)$$ $$\lim_{T\to+\infty} \delta^{a*}(0,q) = \frac 1\gamma \ln\left(1+\frac \gamma k\right) + \frac 1k \ln\left(\frac{f^0_{q}}{f^0_{q-1}}\right)$$

Turning to the characterization of $f^0$ stated in Theorem 2, we just need to write the Rayleigh ratio associated to the smallest eigenvalue of $M+2\eta I$:

$$f^0 \in \underset{f \in \mathbb{R}^{2Q+1}, \|f\|_2 = 1}{\mathrm{argmin}} f'(M+2\eta I) f$$
Equivalently:
$$f^0 \in \underset{f \in \mathbb{R}^{2Q+1}, \|f\|_2 = 1}{\mathrm{argmin}} \sum_{q=-Q}^Q \alpha q^2 {f_q}^2 + \eta \sum_{q=-Q}^{Q-1} (f_{q+1} - f_q)^2 + \eta {f_Q}^2 + \eta {f_{-Q}}^2$$\qed

\textbf{Proof of Proposition 3:}\\

Let us first introduce $H = \lbrace u \in L^1_{loc}(\mathbb{R}) / x\mapsto xu(x) \in L^2(\mathbb{R}) \mathrm{\; and \; } u' \in L^2(\mathbb{R}) \rbrace$.\\ $H$ equipped with the norm $\|u\|_{H} = \sqrt{\int_{\mathbb{R}} \left(\alpha x^2 u(x)^2 + \eta u'(x)^2\right) dx}$ is an Hilbert space.\\

Step 1: $H \subset L^2(\mathbb{R})$ with continuous injection.\\

Let us consider $u \in H$ and $\epsilon > 0$.\\

We have: $$\int_{\mathbb{R}\setminus[-\epsilon,\epsilon]} u(x)^2 dx \le \frac {1}{\epsilon^2}\int_{\mathbb{R}\setminus[-\epsilon,\epsilon]} x^2 u(x)^2 dx < +\infty$$
Hence because $u' \in L^2(\mathbb{R})$, we have $u\in H^1(\mathbb{R}\setminus[-\epsilon,\epsilon])$ with a constant $C_\epsilon$ independent of $u$ such that $\|u\|_{H^1(\mathbb{R}\setminus[-\epsilon,\epsilon])} \le C_\epsilon \|u\|_H$. In particular $u$ is continuous on $\mathbb{R}^*$.\\

Now, if $\epsilon=1$, $\forall x \in (0,1), u(x) = u(1) - \int_x^1 u'(t) dt$ and then $|u(x)| \le |u(1)| + \sqrt{1-x} \|u'\|_{L^2((0,1))}$.\\

Because the injection of $H^1((1,+\infty))$ in $C([1,+\infty))$ is continuous, we know that there exists a constant $C$ independent of $u$ such that $|u(1)| \le C \|u\|_{H^1((1,+\infty))}$. Hence, there exists a constant $C'$ such that $|u(1)| \le C' \|u\|_{H}$ and eventually a constant $C''$ such that $\|u\|_{L^\infty((0,1))} \le C'' \|u\|_H$. Similarly, we obtain $\|u\|_{L^\infty((-1,0))} \le C'' \|u\|_H$.\\

Combining the above inequalities we obtain a new constant $K$ so that $\|u\|_{L^2(\mathbb{R})} \le K \|u\|_H$.\\\qed

A consequence of this first step is that $H \subset H^1(\mathbb{R}) \subset C(\mathbb{R})$.\\

Step 2: The injection $H \hookrightarrow L^2(\mathbb{R})$ is compact.\\

Let us consider a sequence $(u_n)_n$ of functions in $H$ with $\sup_n \|u_n\|_H < +\infty$.\\

Because $H \subset H^1(\mathbb{R})$, $\forall m\in \mathbb{N}^*$, we can extract from $(u_n)_n$ a sequence that converges in $L^2((-m,m))$. Using then a diagonal extraction, there exists a subsequence of $(u_n)_n$, still denoted $(u_n)_n$, and a function $u \in L^2_{loc}(\mathbb{R})$ such that $u_n(x) \to u(x)$ for almost every $x \in\mathbb{R}$ and $u_n \to u$ in the $L^2_{loc}(\mathbb{R})$ sense.

Now, by Fatou's lemma: $$\int_{\mathbb{R}} x^2 u(x)^2 dx  \le \liminf_{n \to \infty} \int_{\mathbb{R}} x^2 u_n(x)^2 dx \le \frac{\sup_n \|u_n\|^2_H}{\alpha}$$

Hence, there exists a constant $C$ such that $\forall m\in \mathbb{N}^*$:

$$ \int_{\mathbb{R}} |u(x)-u_n(x)|^2 dx \le \int_{-m}^m |u(x)-u_n(x)|^2 dx + \frac 1{m^2} \int_{\mathbb{R}\setminus[-m,m]}  x^2 |u(x)-u_n(x)|^2 dx$$
$$\le \int_{-m}^m |u(x)-u_n(x)|^2 dx + \frac C{m^2}$$

Hence $\limsup_{n \to \infty} \int_{\mathbb{R}} |u(x)-u_n(x)|^2 dx \le \frac C{m^2}$.\\

Sending $m$ to $+\infty$ we get:
$$\limsup_{n \to \infty} \int_{\mathbb{R}} |u(x)-u_n(x)|^2 dx = 0$$
Hence $(u_n)_n$ converges towards $u$ in the $L^2(\mathbb{R})$ sense.\\\qed

Now, we consider the equation $-\eta u''(x) + \alpha x^2 u(x) = f(x)$ for $f \in L^2(\mathbb{R})$ and we define $u = Lf$ the weak solution of this equation, \emph{i.e.}:

$$\forall v \in H, \int_{\mathbb{R}} \left(\alpha x^2 u(x)v(x) + \eta u'(x)v'(x)\right) dx = \int_{\mathbb{R}} f(x) v(x) dx$$

Step 3: $L : L^2(\mathbb{R}) \to L^2(\mathbb{R})$ is a well defined linear operator, compact, positive and self-adjoint.\\

For $f \in L^2(\mathbb{R})$, $v \in H \mapsto \int_{\mathbb{R}} f(x) v(x) dx$ is a continuous linear form on $H$ because the injection $H \hookrightarrow L^2(\mathbb{R})$ is continuous. Hence, by Lax-Milgram or Riesz's representation theorem, there exists a unique $u \in H$ weak solution of the above equation and $L$ is a well defined linear operator.\\

Now, $\|Lf\|^2_H = \langle f, Lf \rangle \le \|f\|_{L^2(\mathbb{R})} \|Lf\|_{L^2(\mathbb{R})}$. Hence, since the injection $H \hookrightarrow L^2(\mathbb{R})$ is continuous, there exists a constant $C$ such that $\|Lf\|^2_H \le C \|f\|_{L^2(\mathbb{R})} \|Lf\|_H$, which in turn gives $\|Lf\|_H \le C \|f\|_{L^2(\mathbb{R})}$. Since the injection $H \hookrightarrow L^2(\mathbb{R})$ is compact, we obtain that $L$ is a compact operator.\\

$L$ is a positive operator because $\langle f, Lf \rangle = \|Lf\|^2_H \ge 0$.\\

Eventually, $L$ is self-adjoint because $\forall f,g \in L^2(\mathbb{R})$:
$$\langle f,Lg \rangle = \int_{\mathbb{R}} \left(\alpha x^2 Lf(x)Lg(x) + \eta (Lf)'(x)(Lg)'(x)\right) dx$$$$  = \int_{\mathbb{R}} \left(\alpha x^2 Lg(x)Lf(x) + \eta (Lg)'(x)(Lf)'(x)\right) dx = \langle g,Lf \rangle$$\qed

Now, using the spectral decomposition of $L$ and classical results on Rayleigh ratios we know that the eigenfunctions $f$ corresponding to the largest eigenvalue $\lambda^0$ of $L$ satisfy:

$$\frac 1{\lambda^0} = \frac{\|f\|_H}{\|f\|_{L^2(\mathbb{R})}} = \inf_{g \in H\setminus\lbrace 0 \rbrace} \frac{\|g\|_H}{\|g\|_{L^2(\mathbb{R})}}$$

Hence, our problem boils down to proving that the largest eigenvalue of $L$ is simple and that $g: x \mapsto \exp\left(-\frac{1}{2}\sqrt{\frac{\alpha}{\eta}}x^2\right)$ is an eigenfunction corresponding to this eigenvalue (it is straightforward that $g \in H$).\\

Step 4: Any  positive eigenfunction corresponds to the largest eigenvalue of $L$.\\

By definition of $\|\cdot\|_H$, $\forall f \in H, \frac{\||f|\|_H}{\||f|\|_{L^2(\mathbb{R})}} = \frac{\|f\|_H}{\|f\|_{L^2(\mathbb{R})}}$. Hence, if $f$ is an eigenfunction of $L$ corresponding to the eigenvalue $\lambda^0$, then $|f|$ is also an eigenfunction of $L$ corresponding to the eigenvalue $\lambda^0$. Now, if $\tilde{f}$ is an eigenfunction of $L$ corresponding to an eigenvalue $\lambda \neq \lambda^0$, $\langle |f| , \tilde{f} \rangle = 0$. Therefore $\tilde{f}$ cannot be  positive.\\\qed

Step 5: $g$ spans the eigenspace corresponding to the largest eigenvalue of $L$.\\

Differentiating $g$ twice, we get $g''(x) = -\sqrt{\frac{\alpha}{\eta}} g(x)  + \frac{\alpha}{\eta} x^2 g(x)$.\\ Hence $-\eta g''(x) + \alpha x^2 g(x) = \sqrt{{\alpha}{\eta}} g(x)$ and $g$ is a  positive eigenfunction, necessarily associated to the eigenvalue $\lambda^0$ that is therefore equal to $\frac 1{\sqrt{{\alpha}{\eta}}}$.\\

Now, if we look for an eigenfunction $f \in C^\infty(\mathbb{R}) \cap H$ -- because any eigenfunction of $L$ is in $C^\infty(\mathbb{R})$ -- we can look for $f$ of the form $f=gh$. This gives:

\begin{eqnarray*}
0 &=& -\eta f''(x) + \alpha x^2 f(x) - \sqrt{{\alpha}{\eta}} f(x)\\
 &=& -\eta \left(g''(x)h(x) + 2g'(x)h'(x) + g(x)h''(x)\right) + \alpha x^2 g(x)h(x) - \sqrt{{\alpha}{\eta}} g(x)h(x)
\end{eqnarray*}

Hence:

\begin{eqnarray*}
% \nonumber to remove numbering (before each equation)
  0 &=& 2g'(x)h'(x) + g(x)h''(x) = -2 x \sqrt{\frac{\alpha}{\eta}} g(x) h'(x) + g(x) h''(x) \\
&\Rightarrow& h''(x) = 2 x \sqrt{\frac{\alpha}{\eta}} h'(x)\\
&\Rightarrow& \exists K_1, \quad h'(x) = K_1 \exp\left(\sqrt{\frac{\alpha}{\eta}}x^2\right)\\
&\Rightarrow& \exists K_1, K_2, \quad h(x) = K_1 \int_0^x \exp\left(\sqrt{\frac{\alpha}{\eta}}t^2\right) dt + K_2\\
&\Rightarrow& \exists K_1, K_2, \quad f(x) = K_1 g(x) \int_0^x \exp\left(\sqrt{\frac{\alpha}{\eta}}t^2\right) dt + K_2 g(x)\\
\end{eqnarray*}

Now,

$$g(x) \int_0^x \exp\left(\sqrt{\frac{\alpha}{\eta}}t^2\right) dt \ge \exp\left(-\frac 12 \sqrt{\frac{\alpha}{\eta}}x^2\right) \int_{\frac {x}{\sqrt{2}}}^x \exp\left(\sqrt{\frac{\alpha}{\eta}}t^2\right) dt$$$$ \ge x \left(1 - \frac {1}{\sqrt{2}}\right)$$

Hence, for $f$ to be in $H$, we must have $K_1=0$. Thus, $g$ spans the eigenspace corresponding to the largest eigenvalue of $L$ and Proposition 3 is proved.\\\qed

\end{document}